\documentclass[journal]{IEEEtran}

\ifCLASSINFOpdf
\usepackage[pdftex]{graphicx}
   \graphicspath{{figs/}}
   \DeclareGraphicsExtensions{.pdf,.jpeg,.png}
\else
   \usepackage[dvips]{graphicx}
   \graphicspath{{figs/}}
   \DeclareGraphicsExtensions{.eps}
\fi

\usepackage{longtable}
\usepackage{booktabs} 
\usepackage{caption}
\usepackage{comment}
\usepackage{mathrsfs}
\usepackage{amsmath}
\usepackage{pifont}
\usepackage{amsthm}
\usepackage{algorithm}
\usepackage{algpseudocode}
\usepackage{multirow}
\usepackage{array}
\usepackage{mdwmath}
\usepackage{mdwtab}
\usepackage{eqparbox}
\usepackage{subfig}
\usepackage{threeparttable}
\usepackage{framed}
\usepackage{tabularx}
\usepackage{enumitem}
\usepackage{mathtools}
\usepackage{makecell}
\usepackage{float}
\usepackage{placeins}
\usepackage{graphicx} 	
\usepackage{textcomp}
\usepackage{float}
\usepackage{amssymb}
\usepackage{subcaption}
\usepackage{afterpage}
\usepackage{cite}

\ifCLASSINFOpdf
\else
\fi
\begin{document}
%
\title{Securing Cross-Domain Internet of Drones: An RFF-PUF Allied Authenticated Key Exchange Protocol With Over-the-Air Enrollment}

%
%
%

\author{Xuanyu~Chen,~\IEEEmembership{Student Member,~IEEE},
        Yue~Zheng,~\IEEEmembership{Member,~IEEE},        
        Junqing~Zhang,~\IEEEmembership{Senior Member,~IEEE},
        Guanxiong~Shen,~\IEEEmembership{Member,~IEEE} and
        Chip-Hong~Chang,~\IEEEmembership{Fellow,~IEEE}
\thanks{Manuscript received xxx; revised xxx; accepted xxx. Date of publication xxx; date of current version xxx. The review of this paper was coordinated by xxx.  \textit{(Corresponding author: Yue Zheng.)}}
\thanks{X. ~Chen and Y. ~Zheng are with the School of Science and Engineering, The Chinese University of Hong Kong, Shenzhen, Guangdong, 518172, P.R. China. (emails: xuanyuchen2@link.cuhk.edu.cn, zhengyue@cuhk.edu.cn)}
\thanks{J.~Zhang is with the School of Computer Science and Informatics, University of Liverpool, Liverpool, L69 3DR, United Kingdom. (email: Junqing.Zhang@liverpool.ac.uk)}
\thanks{G. ~Shen is with the School of Cyber Science and Engineering, Southeast University, Nanjing 210096, China (e-mail: gxshen@seu.edu.cn)}
\thanks{C.H. ~Chang is with the School of Electrical and Electronic Engineering, Nanyang Technological University, Singapore 639798 (email: echchang@ntu.edu.sg)}
\thanks{
}
\thanks{
}
}

%
%

\markboth{Journal of \LaTeX\ Class Files,~Vol.~14, No.~8, Month~Year}%
{Shell \MakeLowercase{\textit{et al. }}: Bare Demo of IEEEtran.cls for IEEE Journals}
%



\maketitle

\begin{abstract}

The Internet of Drones (IoD) is an emerging and crucial paradigm enabling advanced applications that require seamless, secure communication across heterogeneous and untrusted domains. In such environments, access control and the transmission of sensitive data pose significant security challenges for IoD systems, necessitating the design of lightweight mutual authentication and key exchange protocols. Existing solutions are often hampered by high computation overhead, reliance on third parties, the requirement for secret storage in resource-constrained drones, and the need for a strictly controlled enrollment environment. These limitations make them impractical for dynamic cross-domain deployment. To address these limitations, we propose a lightweight mutual authentication mechanism that integrates Radio Frequency Fingerprint (RFF) and Physical Unclonable Function (PUF) technologies for secure drone-to-drone (D2D) and drone-to-ground station server (D2G) communication. RFF-based device identification is used to achieve over-the-air (OTA) enrollment, while the PUF serves as the root of trust for establishing mutual authentication among communication parties. Additionally, the on-the-fly key generation capability of the PUF is co-designed with One-Time-Pad (OTP) encryption to realize ephemeral keying and eliminate the need for storing secrets within drones. Both informal security analysis and ProVerif-based formal security verification comprehensively demonstrate the resilience of our protocol against common security attacks. The proposed protocol also outperforms existing IoD authentication schemes in terms of security features, as well as computation, communication, and storage overhead.

\end{abstract}


%
\IEEEpeerreviewmaketitle

\section{Introduction}
\label{sec:introduction}

The Internet of Drones (IoD) refers to a network of interconnected drones, also known as Unmanned Aerial Vehicles (UAVs). These drones communicate with each other and with ground control systems. The IoD offers significant advantages in control operation, flexible deployment, and wide-area sensing. This enables a wide range of applications such as transportation and logistics, security inspections, geographic mapping, emergency rescue, etc.~\cite{li2018uav}. As these applications increasingly demand seamless integration across diverse environments (for example, emergency response drones transitioning between military and civilian airspace, or logistics drones traversing city, regional, and international boundaries), the ability for drones to operate across domains becomes a critical emerging requirement. 

IoD systems typically involve extensive data collection and exchange among network entities. Since this data is often private and sensitive, robust security measures are essential to prevent adversaries from intercepting messages. Access control presents another critical security concern. A successful compromise of a legitimate drone would enable an adversary to not only exfiltrate critical information but also manipulate other drones by transmitting erroneous information. Given the resource and battery constraints of drones, as well as their real-time operational demands, it is essential to design lightweight mutual authentication and key exchange protocols for IoD systems~\cite{mozaffari2019tutorial,mekdad2024comprehensive}.

Numerous authentication and key exchange protocols have been proposed for IoD systems. Many of these solutions rely on traditional cryptography, especially the Elliptic Curve Cryptography (ECC). ECC has become a popular choice because it offers an equivalent security level to other public key cryptosystems while using smaller key sizes. For example, Hussain \textit{et al.}~\cite{hussain2021amassing} proposed an ECC-based three-factor authentication and key agreement (AKA) scheme to secure user-drone communication. However, this scheme is vulnerable to session-key disclosure and UAV impersonation attacks~\cite{wu2021amassing}. Similarly, Ko \textit{et al.}~\cite{ko2021drone} presented an Elliptic Curve Discrete Logarithm Problem-based Mutual Authentication and Key Exchange (MAKE) scheme for both drone-to-drone and drone-to-server communication. This work, however, suffers from very high communication costs, leading to increased power consumption and transmission latency. Subsequently, blockchain-based solutions were introduced to address the dual challenges of non-repudiation and secure information storage. Dwivedi \textit{et al.}~\cite{dwivedi2023d3apts} proposed an ECC-based AKA scheme with blockchain integration for tactile Internet-enabled IoD systems. While robust against multiple attack vectors due to the combined use of ECC and distributed ledger technology, this scheme requires a ground station server (GSS) proxy to achieve mutual authentication between users and drones. Yu \textit{et al.}~\cite{yu2024rlba} proposed RLBA-UAV, a robust blockchain-based AKA scheme for Physical Unclonable Function (PUF)-enabled UAVs. This scheme leverages blockchain to mitigate single-point failures and ensure secure information storage. A common drawback of these latter approaches is the reliance on an intermediate third party as a proxy for authentication, which, like other blockchain-based authentication protocols~\cite{shahidinejad2023anonymous,huang2024bakas}, introduces considerable communication delay. Furthermore, many current blockchain-based protocols often fail to account for the physical security of drones, leaving them susceptible to real-world threats like tampering, capture, and cloning attacks.

PUFs have emerged as a promising lightweight security solution for the IoD. PUFs leverage the inherent, random manufacturing variations in modern semiconductors to generate unique, unclonable device features, often referred to as a device “fingerprint”. This technology is widely utilized for dynamic key generation and device authentication, as demonstrated across various PUF-based IoD authentication protocols~\cite{pu2022lightweight, zhang2025csap, lounis2022d2d, karmakar2023puf}. The rise of PUFs as a lightweight hardware security primitive has driven significant research into developing effective PUF-based authentication protocols for the IoD systems. A review of existing literature highlights several key approaches and their inherent limitations. Pu \textit{et al.}~\cite{pu2022lightweight} adopted PUFs and a chaotic system for lightweight, low-computation mutual authentication between drones. Its primary weakness, however, is the necessity of storing raw Challenge-Response Pairs (CRPs) at the verifier, which makes it vulnerable to stolen verifier attacks. Furthermore, similar to other protocols~\cite{yu2024rlba, huang2024bakas, karmakar2023puf, bansal2024secure, sen2025securing}, it is not communication-optimized because it relies on a ground station for mutual authentication. Lounis \textit{et al.}~\cite{lounis2022d2d} introduced a third-party-free, drone-to-drone mutual authentication protocol. It used the notion of extended CRP to avoid storing plaintext CRPs at the verifier, thereby preventing CRP disclosure attacks. Despite this innovation, the protocol was later shown to be vulnerable to Denial of Service (DoS) attacks. Bansal \textit{et al.}~\cite{bansal2024achieving} proposed an AKA scheme to secure communication between a drone and the GSS. It leverages Shamir's secret sharing to address the reliability concerns inherent to PUFs. The drawback of this approach is that it greatly increases the number of necessary network communications, rendering it unsuitable for UAV swarms operating under limited network bandwidth constraints.

After close scrutiny of existing solutions, we identify the following limitations: 
\begin{enumerate}[leftmargin=*]
\item \textbf{High computational overhead}: Most protocols involve expensive cryptographic operations (e.g., public key encryption and decryption, signature generation and verification, point multiplication, bilinear pairing, etc.), which consume substantial computational resources at drones~\cite{hussain2021amassing,ko2021drone,dwivedi2023d3apts}. 
\item \textbf{Dependence on third party}: Communication between drones often requires intermediaries, inevitably introducing delays and privacy risks~\cite{yu2024rlba,huang2024bakas,pu2022lightweight,sen2025securing, karmakar2023puf}.
\item \textbf{Insecure secret storage}: Many protocols still require secrets to be stored in the drone’s memory, whose security is hard to guarantee due to the limited resources and exposed deployment environment of drones~\cite{lounis2022d2d,karmakar2023puf,zhou2025radio}. 
\item \textbf{Inflexible enrollment constraints for cross-domain operations}: Most protocols necessitate a secure environment for the initial enrollment of drones. This rigidity becomes a critical roadblock in dynamic, cross-domain IoD, where drones frequently transition into new domains and require rapid, repeated re-enrollment.
\end{enumerate}

Limitation 4 is the most significant barrier to achieving dynamic cross-domain IoD. The core conflict lies in the requirement for a secure environment during every enrollment event, which fundamentally hinders the ability of drones to move seamlessly across untrusted domains. Traditional security methods, such as digital signatures~\cite{kaur2012digital} are inadequate because they impose a heavy burden on resource-constrained drones. This burden stems from their high computational demands and the necessity of securely storing sensitive secrets. A new solution is direly needed. This solution must be genuinely lightweight, intrinsically resistant to common threats like replay and Man-in-the-Middle (MITM) attacks, and eliminate the need for secure secret storage or reliance on a protected enrollment environment.

Radio Frequency Fingerprint Identification (RFFI) is emerging as a transformative solution for device identification. Similar to PUFs, RFFI leverages unique device characteristics, but crucially, it eliminates the need for pre-shared secrets. RFFI exploits the intrinsic and unclonable imperfections of a device’s radio-frequency (RF) front-end. These imperfections, arising from manufacturing variations in analog component parameters~\cite{zhang2025physical}, form a unique hardware fingerprint that differentiates each individual transmitter. They subtly distort the emitted radio signals without disrupting normal communication~\cite{sankhe2019no}. By extracting these unique fingerprints from the received waveforms, the receiver can accurately determine the identity of the transmitter. Notably, RFFI offers inherent resistance to common communication security threats, such as replay and MITM attacks~\cite{zhang2022radio}. Furthermore, RFFI imposes no additional computational, storage, or hardware modification burdens on resource-constrained devices like drones in the IoD scenario. Device identification only requires model deployment on a resource-rich GSS. Leveraging these attributes, RFFI presents a strong promising solution to bypass secure enrollment constraints in IoD scenarios.

To date, only one existing authentication protocol~\cite{zhou2025radio} is known to incorporate RFFI. This scheme achieves AKA between edge devices and a server. Unlike conventional RFFI implementations, this protocol integrates RF fingerprints into a multi-factor authentication framework by deriving a secret string from the physical-layer characteristics for cryptographic operations. However, the protocol has a critical limitation: edge devices must store secrets locally. This requirement poses a significant security concern regarding physical attacks and potential compromise if a device (e.g., a drone) is lost, stolen, or crash lands in an unsecured location. Crucially, this design choice fails to capitalize on RFFI's core benefit: the elimination of pre-shared secrets and on-device cryptographic overhead, particularly during cross-domain enrollment.

In this paper, we propose a novel dual root-of-trust framework that integrates PUF and RFFI to enable secure MAKE for IoD systems. Our protocol consists two main phases. In the enrollment phase, the RFFI serves as the initial root of trust, ensuring secure, over-the-air (OTA) provisioning without requiring a trusted environment or pre-shared secrets. Subsequently, in the MAKE phase, the PUF acts as the root of trust, guaranteeing direct, lightweight, and key-storage-free authentication directly between drones. Our main contributions are: 
\begin{itemize}
\item \textbf{Secure OTA Enrollment:} We introduce a wireless provisioning paradigm that uses RFFI and ECC-based public key cryptography (PKC) to eliminate physical access requirements. Importantly, devices do not need transmitter modification or resource-intensive ECC decryption, enabling fully remote deployment in open environments and significantly enhancing applicability and scalability.  

\item \textbf{Direct and Secure MAKE:} Our scheme achieves direct and secure authentication and key exchange between drones without relying on a trusted third party. By leveraging the PUF for on-the-fly key generation and the One-Time Pad (OTP) for cryptographic concealment, we eliminate the need for secure secret storage on devices, achieving enhanced security and hardware efficiency.

\item \textbf{Enhanced Security Properties and Efficiency:} Perfect forward secrecy (PFS) and replay attack resistance are achieved by dynamically updating the PUF-based long-term keys during protocol execution. Furthermore, communication overhead is significantly reduced by accomplishing the entire MAKE with only two message exchanges.

\item \textbf{Rigorous Analysis and Verification:} The proposed protocol has undergone both informal analysis and automatic verification using the professional security verification tool, ProVerif~\cite{ProVerifManual}. We also present a comprehensive comparison of our proposed scheme with existing IoD solutions, evaluating security features, computation complexity, communication cost, and secret storage requirements.

\end{itemize}

The rest of the paper is organized as follows. Section~\ref{sec:preliminaries} provides preliminaries. Section \ref{sec:system} describes the system model and security requirements. In Section~\ref{sec:protocol}, we present the proposed protocol design, detailing the secure registration, OTA enrollment, and MAKE phases. We conduct the security analysis in Section~\ref{sec:security} and present the performance comparisons in Section~\ref{sec:performance}. Finally, Section~\ref{sec:conclusion} concludes the paper.

\section{Preliminaries}
\label{sec:preliminaries}
\subsection{PUF}
\label{sec:puf}
A PUF~\cite{pappu2002physical, chang2017retrospective} is a physical system that implements a unique and unclonable mapping from a challenge space $\mathcal{C}$ to a response space $\mathcal{R}$, such that $ R = \emph{PUF}(C)$ where $C\leftarrow\mathcal{C}$ and $R\leftarrow\mathcal{R}$. Each such $(C, R)$ pair is called a CRP. Due to random manufacturing variations, it is computationally infeasible to reproduce its CRPs without physical possession of the device. The following properties are essential for a robustly designed PUF to be cryptographically useful:
\begin{itemize}[leftmargin=*]
\item \textit{Evaluability}: Given a PUF device and a challenge $C\leftarrow\mathcal{C}$, the corresponding response $R\leftarrow\mathcal{R}$ should be generated efficiently.
    \item \textit{One-Wayness}: A PUF instance $\emph{PUF}_i$ is said to be akin to a physical one-way function if it is computationally and physically infeasible to clone or reproduce a device to exhibit the same challenge-response behavior by measuring or tampering with the device, even for the original manufacturer. 
    \item \textit{Unpredictability}: It is computationally infeasible for a probabilistic polynomial time~(PPT) adversary to predict the response $R = \emph{PUF}_i(C)$ of any PUF instance $\emph{PUF}_i$ to an unobserved challenge $C$ with a non-negligible probability, even with access to a large number of observed CRPs.
    \item \textit{Uniqueness}: For any given challenge $C$, the responses from two different PUF instances, $\emph{PUF}_1$ and $\emph{PUF}_2$, should be significantly different. This property, measured by the inter-device Hamming distance, approaches 50\% for an ideal PUF. 
    \item \textit{Reliability}: The response $R$ of a PUF instance to any challenge $C$ must be reproducible over time and under various environmental conditions with a very high probability.
\end{itemize}

PUF can be implemented flexibly using either dedicatedly designed circuits, like symmetric combinational circuit delay paths or ring oscillators, or existing circuit components, such as memories or image sensors. Any physical attempt to tamper with a PUF circuit is intended to alter its characteristic, thereby providing evidence of a security breach. However, sophisticated attackers have developed methods that can bypass or exploit PUF vulnerabilities without necessarily destroying its unique identity outright. These vulnerabilities include machine-learning and side-channel attacks, which can model the PUF's behavior and compromise its security. Therefore, robust security protocols are required to manage the PUF's CRPs. For example, it is imperative to carefully manage the CRPs used, such as employing them only once or protecting the communication channel, to prevent eavesdropping, replay, or modeling attacks. 

\subsection{RFFI} 
\label{sec:RFFI}
Leveraging RFF for transmitter identification eliminates the need for pre-shared keys and hardware modifications on the transmitter side. In addition, RFFI is intrinsically secure against replay and MITM attacks. This is because RFFI inherently verifies the physical source of received signals, and any re-transmission would be easily detected due to the mismatch in the physical characteristics of the transmitting device. Owing to these appealing features, RFFI has received increasing attention, with deep learning (DL) being prevalently used for RFFI due to its exceptional feature extraction capabilities. However, most DL-based RFFI schemes adopt a closed-set setting, which is inherently limited to known devices and struggles to identify unknown or adversarial transmitters, posing a critical challenge in dynamic IoD environments~\cite{mohanti2020airid,rajendran2022rf,shen2021radio}.

Recent advancements have introduced more sophisticated open-set RFFI schemes to address the challenge of identifying unknown devices. Unlike closed-set approaches, where a DL model serves as both the feature extractor and classifier, open-set schemes typically decouple RFFI into two components: a DL-based RFF extractor and a classifier. The extractor can be implemented utilizing methods such as an AutoEncoder~\cite{wang2024design}, a ResNet~\cite{xie2025novel}, or a Convolutional Neural Network~\cite{shen2022towards}. Commonly used classifiers include $k$-nearest neighbor (KNN) based on Euclidean distance~\cite{shen2022towards}, distance-to-centroid classification based on Mahalanobis distance~\cite{xie2025novel}, and classification based on cosine similarity~\cite{wang2024design}. 

\begin{figure}[!t]
    \centering
    \includegraphics[width=1\linewidth]{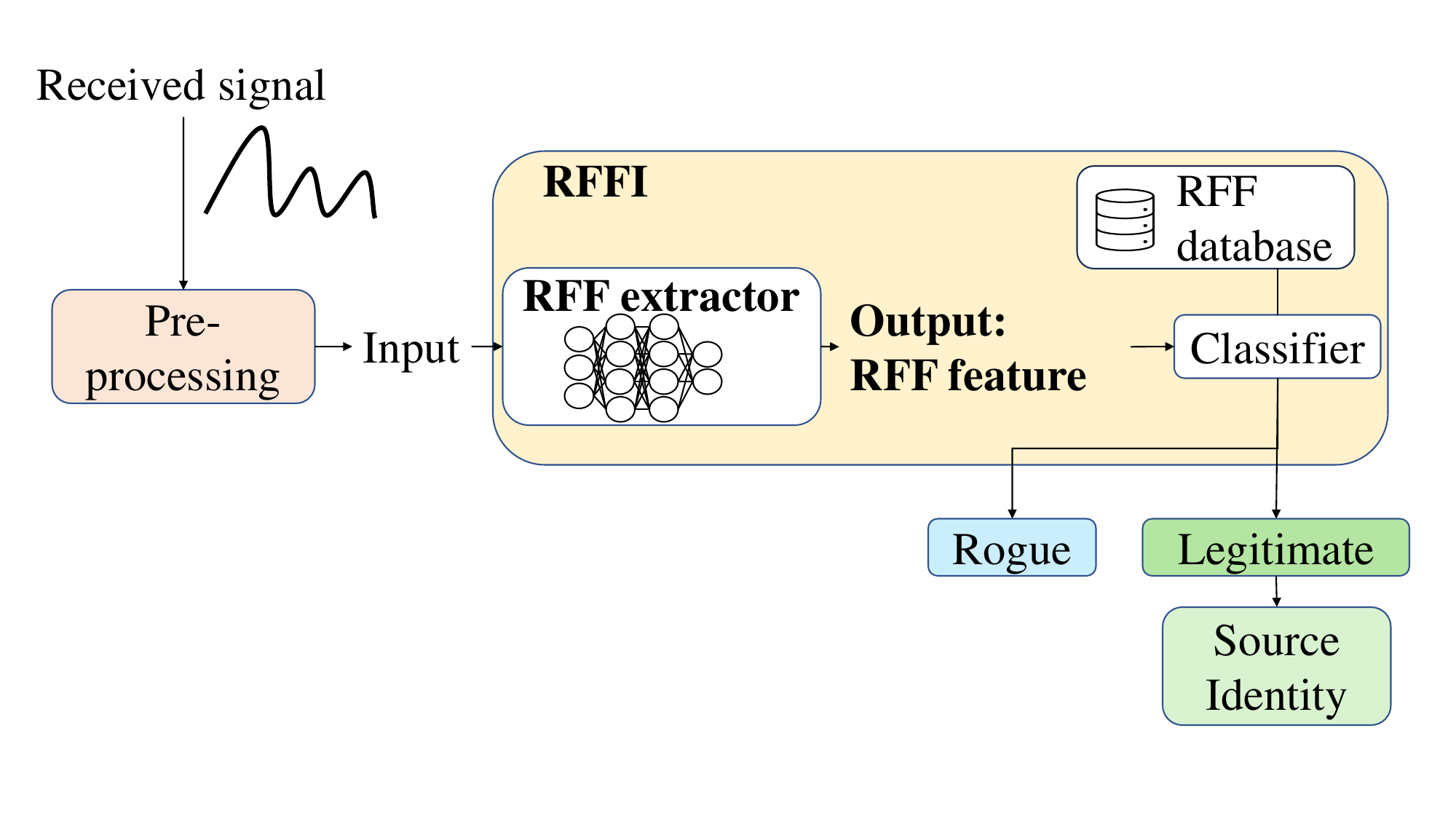}
    \caption{Open-set DL-based RFFI.}
    \label{fig:RFFI}
\end{figure}

The workflow of the open-set RFFI algorithm is depicted in Fig. \ref{fig:RFFI}. Initially, essential signal pre-processing steps, including synchronization, preamble extraction, carrier frequency offset (CFO) compensation, and power normalization, are performed to eliminate bias, instability, and other undesirable components from the received signals. Subsequently, the pre-processed signal is fed into the RFF extractor to generate an RFF feature. This feature is then compared against an RFF database collected during the enrollment phase. The wireless packet is classified at the waveform level to determine its source identity only if the measured similarity, using a metric such as cosine, Mahalanobis, or Euclidean distance, exceeds a predefined threshold. Otherwise, the wireless packet is flagged as malicious and rejected. Recent studies have demonstrated that open-set RFFI schemes achieve strong discriminative performance for rogue device detection, with one approach attaining 0.99 AUC (Area Under the receiver operating characteristic Curve) on 15 devices (10 known legitimate devices and 5 unknown rogue devices)~\cite{shen2022towards} and another achieving 89–97\% accuracy across 0–30 dB SNR (Signal-to-Noise Ratio) with 18 devices (12 known legitimate devices and 5 unknown rogue devices.)~\cite{wang2024design}. Such schemes are also highly scalable because the addition or removal of devices only requires an update to the RFF database. 

\section{System model and security requirements}
\label{sec:system}

\subsection{System Model}
\begin{figure}[!t]
    \centering
    \includegraphics[width=1\linewidth]{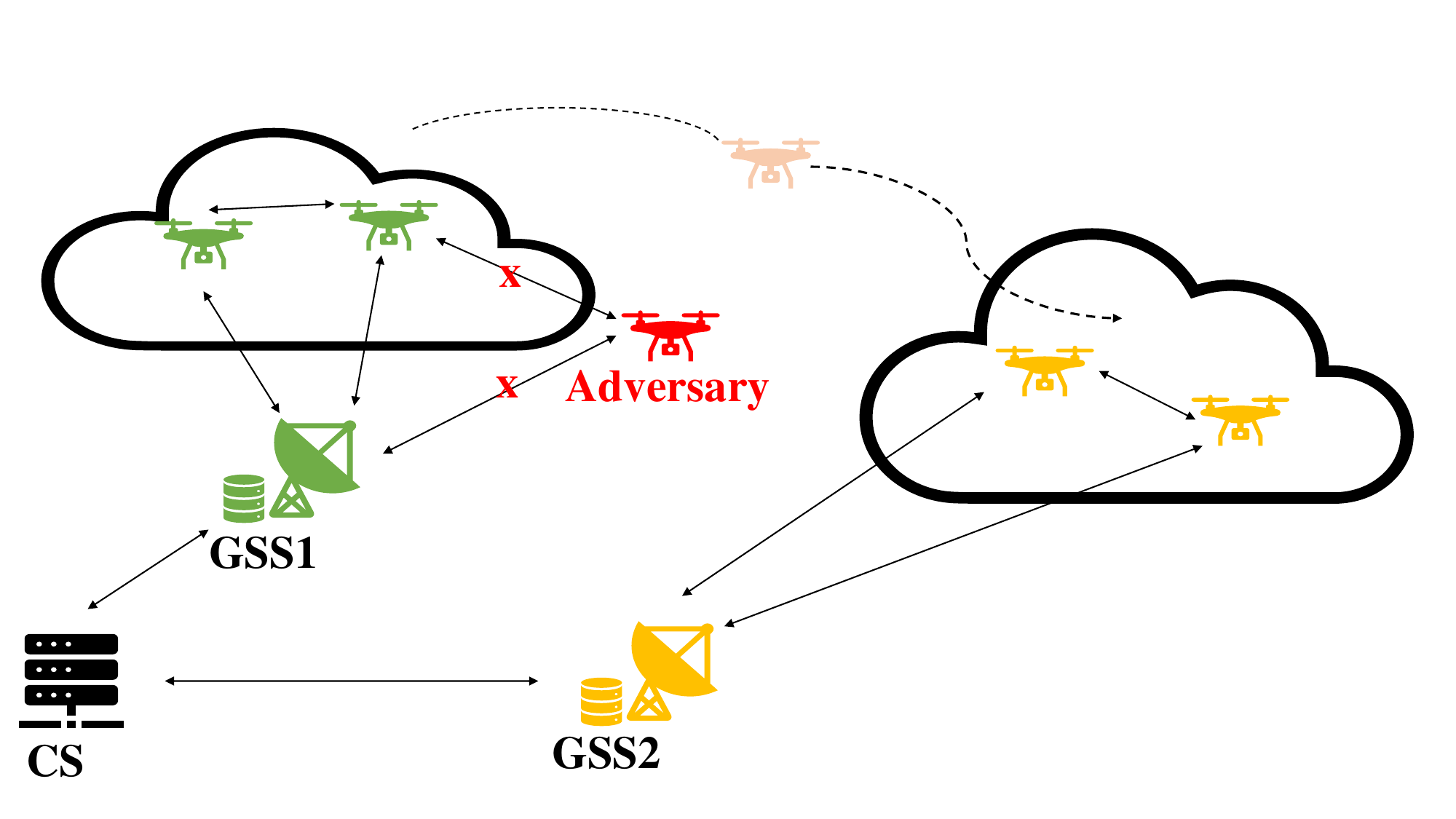}
    \caption{IoD system model.}
    \label{fig:system_model}
\end{figure}

Fig. \ref{fig:system_model} illustrates the system model, comprising three primary entities: the Control or Cloud Server (\texttt{CS}), Ground Station Servers (\texttt{GSSs}), and drones. The \texttt{CS} is a secure, central institutional server primarily responsible for registering drones and \texttt{GSS}s, managing keys, and maintaining critical information about registered devices. It is assumed to be an absolutely credible and controllable part of the overall IoD system infrastructure. \texttt{GSSs} are deployed on the ground and act as a bridge between the drones and the \texttt{CS}. Each \texttt{GSS} oversees a specific domain (area) and controls the drones within its domain. This includes authenticating drones, collecting data from them for processing, and relaying the data to the \texttt{CS}. Drones are mobile nodes within the IoD with limited computational capability and communication bandwidth. They can freely move through different domains to collect data and communicate with the \texttt{GSS} and other drones operating in the same domain. As such, drones can join or leave any domain at any time, forming a dynamic group that requires continuous renewal and revocation of security credentials for its changing membership. We consider two types of communications: drone-to-drone (D2D) and drone-to-\texttt{GSS} (D2G). Drones and the \texttt{GSS} may communicate with each other for tasks such as navigation, coordination, or remote control. Additionally, all \texttt{GSS}s and drones in our system are equipped with PUFs. To detect unknown adversary devices, our scheme leverages the open-set RFFI, as shown in Fig. \ref{fig:RFFI}~\cite{shen2022towards}. A well-trained RFF extractor is assumed to be available at the \texttt{CS}. 

\subsection{Security Requirements} To ensure secure communication in both D2D and D2G scenarios, the proposed protocols should achieve the following security requirements: 

\begin{enumerate}[leftmargin=*]
    \item Mutual Authentication: It is important to guarantee that only authorized entities participate in both D2G and D2D communications. In either scenario, both drones and the \texttt{GSS} should be able to confirm the authenticity and integrity of the received message, as well as the identity of the message’s sender.
    \item Session Key Agreement: To prevent sensitive data from leaking via eavesdropping on public channels, fresh session keys must be generated for secure data exchange between communicating parties in each session.
    \item Perfect Forward Secrecy (PFS): PFS guarantees that previous communications encrypted with ephemeral session keys cannot be decrypted retroactively. This is true even if an adversary gains access to the system’s long-term secrets. The protocol must achieve PFS to guarantee data security.

    \item Resistance to Known Attacks: The protocol should be secure against known common malicious attacks, such as eavesdropping, replay, MITM, and DoS attacks. 
\end{enumerate}

\section{Proposed protocol}
\label{sec:protocol}

This section presents our proposed protocol, which consists of three phases: Registration phase, Enrollment phase, and MAKE phase. Table \ref{tab:notation} lists the notations used in our protocol.

\begin{table}[!t]
\centering
\caption{Notations Used by the Proposed Protocol}
\begin{tabularx}{\linewidth}{@{} lX @{}}
\toprule
\textbf{Notation} & \textbf{Description} \\ \midrule
$R=\emph{PUF}_i(C)$ & The response $R$ obtained by applying the challenge $C$ to the PUF of drone $i$. \\
$\mathcal{H}(X, Y)$ & A hash function with two inputs, $X$ and $Y$. \\
$a \oplus b$ & Exclusive OR of variables $a$ and $b$. \\
$x \overset{R}{\leftarrow} X$ & $x$ is randomly chosen from the set $X$. \\
$\{0,1\}^l$ & The set of all possible bit strings of length $l$. \\ 
$E = \emph{Enc}(P, \emph{pk})$ & PKC encryption: the plaintext $P$ is encrypted with the public key $pk$ to obtain the ciphertext $E$. \\
$P = \emph{Dec}(E,\emph{sk})$ & PKC decryption: the ciphertext $E$ is decrypted with the private key $\emph{sk}$ to obtain the plaintext $P$. \\
$M_i$ & The $i$th message in the D2G enrollment protocol or the D2D MAKE protocol. \\
$ \emph{RFFI}(M)$ & An open-set RFFI for identifying known, authorized drones and rejecting unknown, unauthorized drones based on the RF signal carrying the received message $M$. This function outputs $\emph{ID}_i$ for a legitimate drone $i$ or the label `Rogue’ for any unidentified or rogue devices.\\
$\emph{DB}$ & An RFF database storing legitimate drones' identities and their respective RFF features.\\ 
$\emph{pk}_G$ & Public key of the \texttt{GSS}. \\
$\emph{sk}_G$ & Private key of the \texttt{GSS}. \\ 
$\emph{sk}$ & Session key. \\
$s_\emph{AG}$ & A D2G secret derived from the PUF response of Drone \texttt{A} and the target \texttt{GSS}’s identity. \\
$E_\emph{AG}$ & The ciphertext of $s_\emph{AG}$ encrypted with $\emph{pk}_G$. \\
$s_\emph{BA}$ & A D2D secret dervied from the PUF response of Drone \texttt{B} and the identity of Drone \texttt{A}. \\ 
$\kappa_\emph{AB}$ & An OTP key derived from $s_\emph{AG}$ and the identity of Drone \texttt{B}. \\
$X_\emph{BA}$ & The ciphertext of $s_\emph{BA}$ encrypted with the OTP key $\kappa_\emph{AB}$. \\ 
$X_\emph{AG}$ & The ciphertext of $s_\emph{AG}$ encrypted with the OTP key derived from the PUF response of the \texttt{GSS} and the identity of Drone \texttt{A}. \\ 
$\mathscr{C}_i$ & A credential used to authenticate Drone $i$ in the D2D MAKE phase. \\ 
$\mathscr{C}_G$ & A credential used to authenticate the \texttt{GSS} in the D2G enrollment and MAKE phases. \\ 
\bottomrule
\end{tabularx}
\label{tab:notation}
\end{table}

\subsection{Secure Registration in the \texttt{CS}} 
The registration phase aims to register drones and \texttt{GSSs} with the \texttt{CS} prior to system deployment, and this process must be conducted in a secure environment. Each drone is required to send a hundred packets to the \texttt{CS} for registration. These packets will be preprocessed as described in Section~\ref{sec:RFFI}, and sent to the RFF extractor. Note that the RFFI extracts the unique identification information from device-specific hardware variations in the physical layer of the RF signal. This identification method is independent of the transmitted message’s content (the payload), so the message payload remains available to carry auxiliary information such as the drone’s identity, flight data, or other communication. The \texttt{CS} needs to manage a secure database $\emph{DB}$, storing the identity and a set of extracted RFF features for each registered drone. 

To register a \texttt{GSS}, the \texttt{CS} first generates a unique identity $\emph{ID}_G$ and a public-private key pair $\{\emph{pk}_G, \emph{sk}_G\}$ for it using a PKC system, where $\emph{pk}_G$ and $\emph{sk}_G$ denote the public and private keys, respectively. For lightweight considerations, ECC is a good choice. This key pair, along with the components for the RFFI process -- specifically, the RFF extractor and the drone RFF database $\emph{DB}$ -- is then sent to the \texttt{GSS} through a secure channel. The function $\emph{RFFI}(\cdot)$ encapsulates the entire RFFI process, which includes: signal preprocessing, input transformation, RFF feature extraction using the RFF extractor, rogue detection, and device identification leveraging Euclidean-based $K$-NN with the help of $\emph{DB}$. The \texttt{GSS} uses this function, which outputs a ``Rogue” status for adversary devices or the source identity for legitimate drones, to directly identify all registered drones from their received radio signals during D2G communication without the \texttt{CS}’s involvement. The identity $\emph{ID}_G$ and public key $\emph{pk}_G$ of at least one target \texttt{GSS} for each drone is also sent to the drone and stored in its read-only memory. 

\subsection{Over-The-Air Enrollment}
\label{sec:OTA enrolment}
\begin{figure}[!t]
    \centering
    \includegraphics[width=1\linewidth]{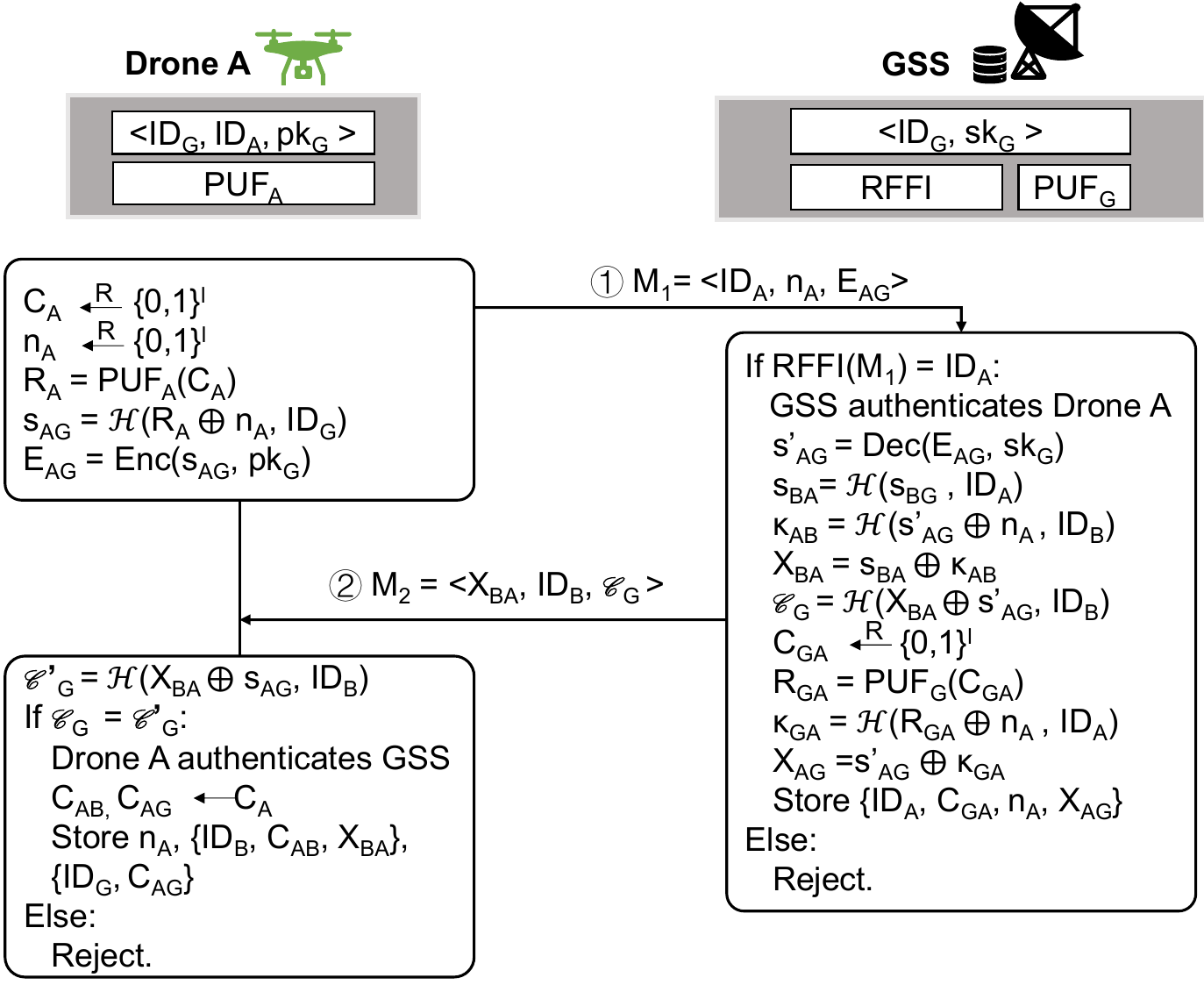}
    \caption{The OTA Enrollment phase.}
    \label{fig:enrollment_GSS}
\end{figure}

Each drone, upon successful registration with the \texttt{CS}, will have the identity $\emph{ID}_G$ and the public key $\emph{pk}_G$ of the target \texttt{GSS} in addition to its own identity for secure communication with the target \texttt{GSS}. Before a newly joined drone can directly and securely communicate with any other enrolled drones in the domain, it needs to enroll with the \texttt{GSS} to obtain the authentication secrets. This enrollment process is performed OTA in an open environment within the domain. The protocol for enrolling a Drone \texttt{A} that newly joins the domain is shown in Fig. \ref{fig:enrollment_GSS}. The enrollment process is elaborated as follows.

\begin{enumerate}[leftmargin=*,label=(\arabic*)] 
\item \textbf{Drone \texttt{A} initiates an admission request}: \\
\makebox[1em]{}To initiate a joining request, Drone \texttt{A} generates a random challenge $C_A$ and a session nonce $n_{A}$. It applies $C_A$ to its embedded PUF to generate the response $R_{A} = \emph{PUF}_A(C_A)$. To prove to the \texttt{GSS} that it has been officially registered with the \texttt{CS} to operate in this domain, and to allow it to authenticate the \texttt{GSS}, Drone \texttt{A} hashes the PUF response $R_A$, the session nonce $n_A$, and the \texttt{GSS}’s identity into a unique D2G secret, $s_\emph{AG}=\mathcal{H}(R_{A} \oplus n_{A}, \emph{ID}_G)$, where $\mathcal{H}(\cdot, \cdot)$ is a multi-input cryptographic hash function, with the input arguments separated by commas. The inputs are first encoded by concatenation with length prepending into a single unambiguous string before being processed by a standard single-input hash function. To ensure that only the legitimate \texttt{GSS} can recover this D2G secret, $s_\emph{AG}$ is ECC-encrypted to the ciphertext $E_\emph{AG}$ using \texttt{GSS}’s public key $\emph{pk}_G$ and then packed with its identity $\emph{ID}_A$, nonce $n_{A}$ into $M_1 = \langle \emph{ID}_A, n_{A}, E_\emph{AG} \rangle$ before it is sent to the \texttt{GSS} over the insecure channel. 

\item \textbf{The \texttt{GSS} processes Drone \texttt{A}’s admission request}: \\
\makebox[1em]{}Upon receiving $M_1$, the \texttt{GSS} first uses the $\emph{RFFI}$ to analyze the waveform of $M_1$ to determine its sender’s identity. Only if $\emph{RFFI}(M_1)$ matches $\emph{ID}_A$ in the payload does the \texttt{GSS} accept Drone \texttt{A} and decrypt $E_\emph{AG}$ by its private key $\emph{sk}_G$ to obtain the D2G secret, $s^\prime_\emph{AG}$\footnote{The superscript $\prime$ is used to differentiate the recovered secret from the originally generated secret.}. \\

\makebox[1em]{}To enable existing drones in the domain to communicate with the newly admitted Drone \texttt{A}, a shared secret is generated between Drone \texttt{A} and any other enrolled drones in this domain using the latter’s D2G secret. For ease of exposition, let Drone \texttt{B} be an existing drone that has been successfully enrolled with a D2G secret $s_\emph{BG}$. The shared secret between Drones \texttt{A} and \texttt{B} can be calculated as $s_\emph{BA}=\mathcal{H}(s_\emph{BG}, \emph{ID}_A)$. 
This secret is then OTP-encypted into $X_\emph{BA}=s_\emph{BA} \oplus \kappa_\emph{AB}$, with $\kappa_\emph{AB}=\mathcal{H}(s^\prime_\emph{AG} \oplus n_A, \emph{ID}_B)$ as the OTP key. $X_\emph{BA}$ can only be decrypted by Drone \texttt{A} as $s_\emph{AG}$ needed to produce $\kappa_\emph{AB}$ can only be generated with $\emph{PUF}_A$. The \texttt{GSS} protects the confidentiality of $s^\prime_\emph{AG}$ and the integrity of $X_\emph{BA}$ and $\emph{ID}_B$ by generating the credential, $\mathscr{C}_G=\mathcal{H}(X_\emph{BA} \oplus s^\prime_\emph{AG}, \emph{ID}_B)$, before they are transmitted as $M_2=\langle X_\emph{BA}, \emph{ID}_B, \mathscr{C}_G \rangle$ to Drone \texttt{A}. \\

\makebox[1em]{}Besides, a random challenge $C_\emph{GA}$ is selected and stored for communication between the \texttt{GSS} and Drone \texttt{A}. This challenge is used to generate the response $R_\emph{GA}$ from the \texttt{GSS}’s PUF, which in turn is used to generate an OTP key $\kappa_\emph{GA}=\mathcal{H}(R_\emph{GA} \oplus n_{A}, \emph{ID}_A)$ to encrypt the D2G secret $s^\prime_\emph{AG}$ into $X_\emph{AG}$ for secure storage. Finally, the \texttt{GSS} stores $\{\emph{ID}_A, C_\emph{GA}, n_\emph{A}, X_\emph{AG}\}$ for the enrolled Drone \texttt{A}. 

\item \textbf{Drone \texttt{A} authenticates the \texttt{GSS} and completes the enrollment process}: \\
\makebox[1em]{}When Drone \texttt{A} receives $M_2$, it computes $\mathscr{C}^\prime_G=\mathcal{H}(X_\emph{BA} \oplus s_\emph{AG}, \emph{ID}_B)$. If $\mathscr{C}^\prime_G=\mathscr{C}_G$, Drone \texttt{A} authenticates the \texttt{GSS} since only the latter has the correct secret key $\emph{sk}_G$ to successfully decrypt $E_\emph{AG}$ and obtain the D2G secret $s_\emph{AG}$. Once authenticated, Drone \texttt{A} assigns $C_A$ to both $C_\emph{AB}$ and $C_\emph{AG}$ and then saves $n_A$, $\{\emph{ID}_B, C_\emph{AB}, X_\emph{BA}\}$ and $\{\emph{ID}_G, C_\emph{AG}\}$ into its memory. $C_\emph{AB}$ and $X_\emph{BA}$ are used by Drone \texttt{A} for mutual authentication with Drone \texttt{B} in the D2D communication phase, while $C_\emph{AG}$ is used by Drone \texttt{A} for mutual authentication with the \texttt{GSS} in the D2G communication phase.
\end{enumerate}

The aforementioned OTA enrollment scheme provides the authorization, access control, regulatory oversight, and tracking necessary for secure communications. It also ensures seamless handover for the dynamic entry and exit of drones within a domain, as the entire deployment process -- especially enrollment -- can be carried out in an open environment. Each legitimate drone must first register with the \texttt{CS} in a secure environment to obtain the public key(s) of the \texttt{GSS}(s) for those domains it is authorized to fly. When a registered drone requests to join a domain, the \texttt{CS} can propagate the drone's RFF feature to the corresponding \texttt{GSS} via a secure connection; this secure transmission can be fulfilled by existing secure transmission protocols between resource-rich entities~\cite{ITU-T_X.800}. To exit a domain, the leaving drone notifies its associated \texttt{GSS} through an established secure communication session. In response, the \texttt{GSS} informs all other drones in the domain to delete the leaving drone’s related information. Building upon the above framework, drones can further achieve seamless cross-domain mobility. This includes requesting the current domain's \texttt{GSS} to relay the \texttt{GSS}’s public key of the next destination, which the drone is authorized to access from the \texttt{CS}, without having to store the public keys of all authorized domains onboard.

\subsection{Mutual Authentication and Key Exchange (MAKE)}
A lightweight MAKE can be performed between any two successfully enrolled drones (D2D) or an enrolled drone and the \texttt{GSS} (D2G) in the same domain. 

\subsubsection{D2D MAKE}
\begin{figure}[!t]
\centering
\subfloat[]{
 \includegraphics[width=1\columnwidth]{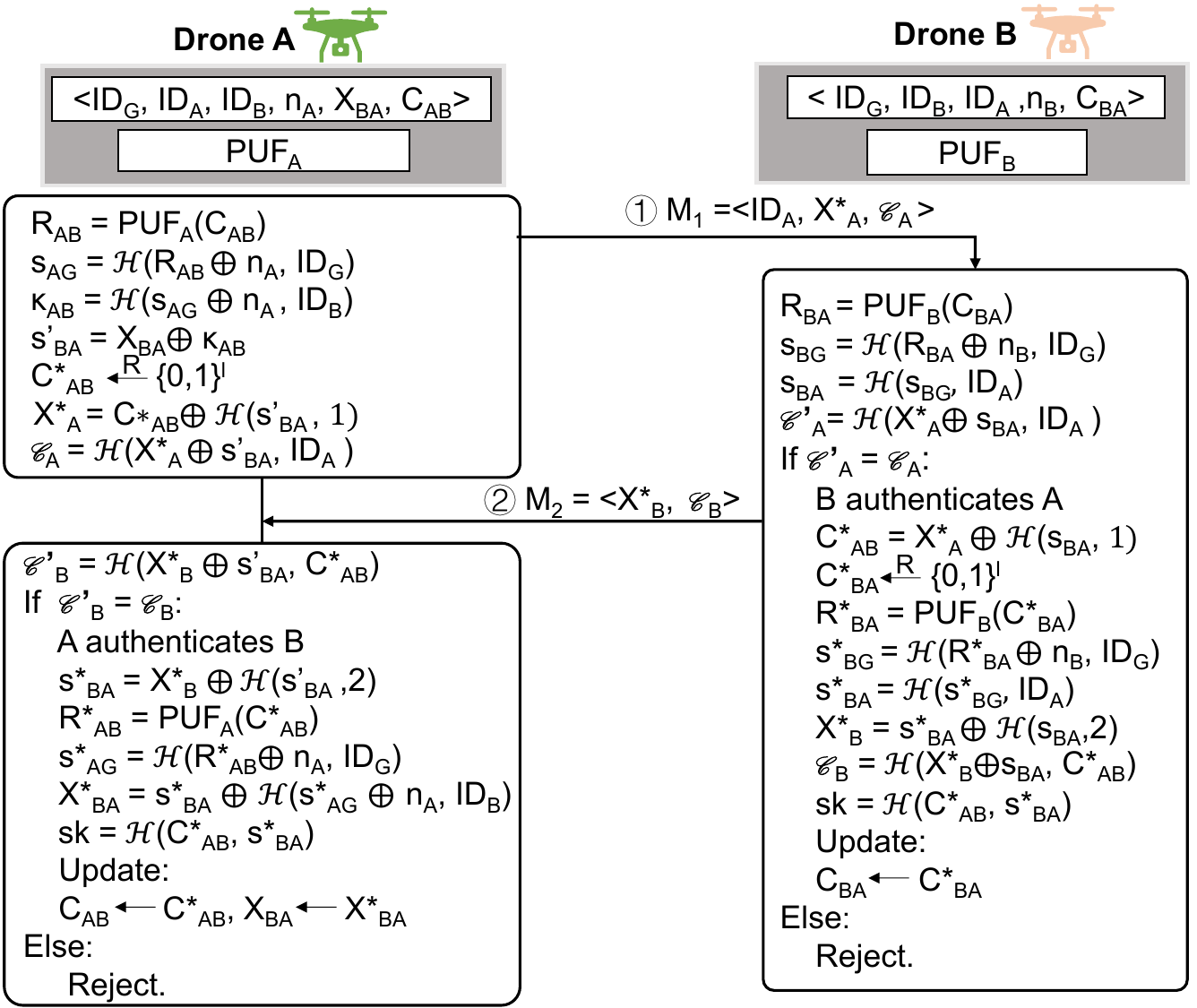}
 \label{fig:D2D_A_initiate}
}
\hfil
\subfloat[]{
 \includegraphics[width=1\columnwidth]{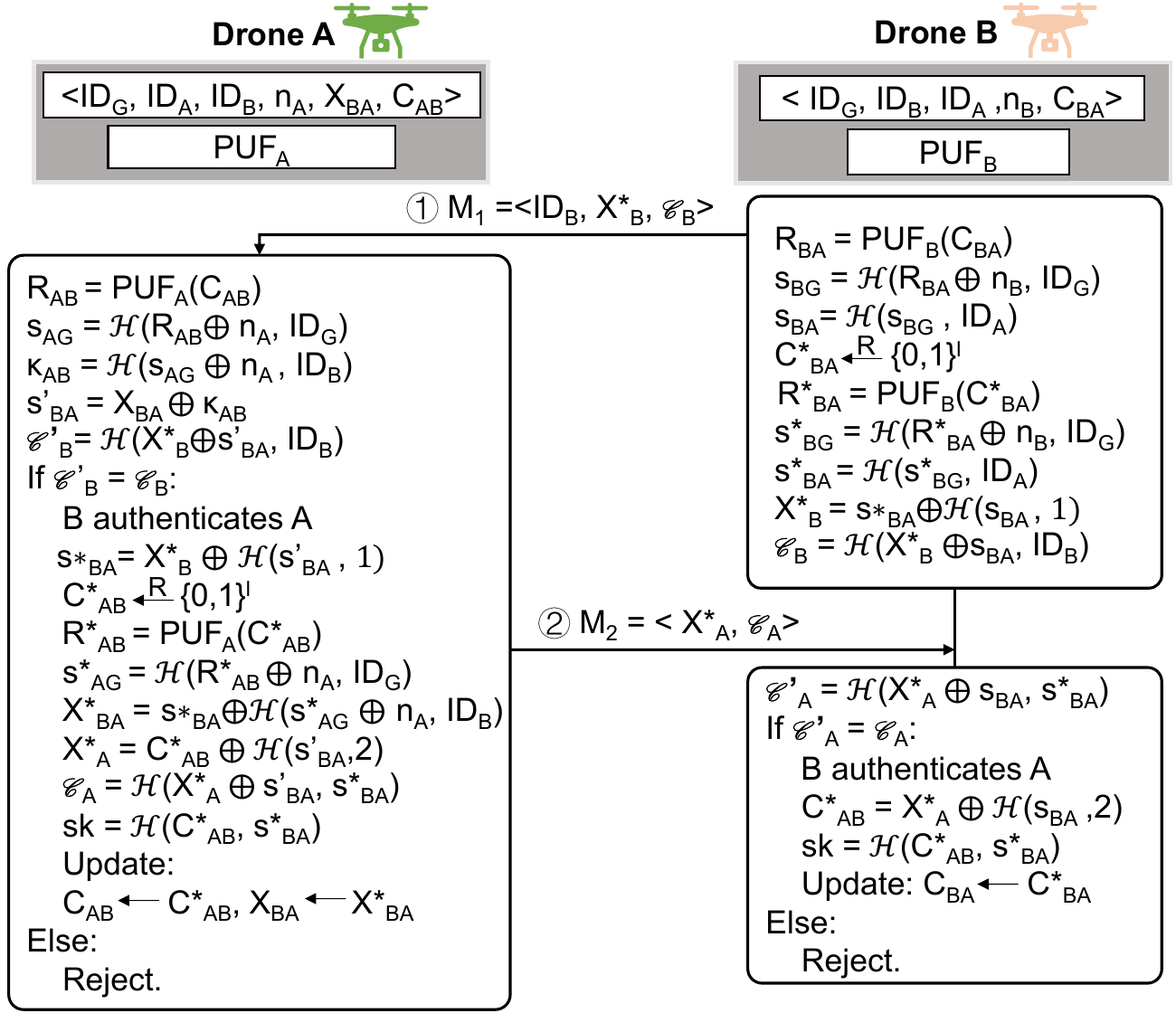}
 \label{fig:D2D_B_initiate}
}
\caption{D2D Mutual Authentication and Key Exchange phase. (a) Initiated by Drone \texttt{A}; (b) Initiated by Drone \texttt{B}.}
\label{fig:D2D} 
\end{figure}

If Drone \texttt{A} is successfully enrolled with the same \texttt{GSS} after Drone \texttt{B}, then Drone \texttt{A} possesses the OTP-encrypted shared secret $X_\emph{BA}$ to authenticate Drone \texttt{B}. Fig. \ref{fig:D2D}(a) presents the D2D MAKE protocol initiated by Drone $A$. The process is elaborated as follows: 
\begin{enumerate}[leftmargin=*,label=(\arabic*)]
    \item \textbf{Drone \texttt{A} recovers the shared secret for generating the authentication credential}:\\
    \makebox[1em]{}To recover the shared D2D secret $s_\emph{BA}$ generated by Drone \texttt{B} from $X_\emph{BA}$, Drone \texttt{A} retrieves $C_\emph{AB}$ and $n_\emph{A}$ from its memory to generate the response $R_\emph{AB} = \emph{PUF}_A(C_\emph{AB})$ and compute $s_\emph{AG}=\mathcal{H}(R_\emph{AB} \oplus n_A, \emph{ID}_G)$. Then, Drone \texttt{A} recovers the shared secret $s^\prime_\emph{BA}$ by XORing $X_\emph{BA}$ with the OTP key $\kappa_\emph{AB}=\mathcal{H}(s_\emph{AG} \oplus n_A, \emph{ID}_B)$. \\
    \makebox[1em]{}The session key \emph{sk} is to be derived from two elements upon successful authentication. One is a new random challenge $C_\emph{AB}^*$\footnote{The superscript asterisk (*) is used to denote an updated variable henceforth.} generated by Drone \texttt{A}. The other is an updated D2D secret generated by Drone \texttt{B}. For Drone \texttt{B} to recover $C_\emph{AB}^*$, this challenge is first masked by $X^*_A=C_\emph{AB}^* \oplus \mathcal{H}(s^\prime_\emph{BA},1)$ and a credential $\mathscr{C}_A = \mathcal{H}(X^*_A \oplus s^\prime_\emph{BA}, \emph{ID}_A)$ is built from it for Drone \texttt{B} to authenticate Drone \texttt{A}. Then, Drone \texttt{A} sends the packet $M_1 = \langle \emph{ID}_A, X^*_A, \mathscr{C}_A \rangle$ to Drone~\texttt{B}.
    
    \item \textbf{Drone \texttt{B} authenticates \texttt{A}, update long-term secrets and reciprocates the authentication credential}:\\
    \makebox[1em]{}Upon receiving $M_1$, Drone \texttt{B} retrieves $C_\emph{BA}$, $n_B$, and $\emph{ID}_G$ from its memory to compute $s_\emph{BG}=\mathcal{H}(R_\emph{BA} \oplus n_B, \emph{ID}_G)$ and the shared secret $s_\emph{BA}=\mathcal{H}(s_\emph{BG}, \emph{ID}_A)$. Then it calculates a credential $\mathscr{C}^\prime_A = \mathcal{H}(X^*_A \oplus s_\emph{BA}, \emph{ID}_A)$. If $\mathscr{C}^\prime_A = \mathscr{C}_A$, Drone \texttt{A} is authenticated, since $s^\prime_\emph{BA}$ recovered by Drone \texttt{A} matches $s_\emph{BA}$ generated by Drone \texttt{B}. \\
    \makebox[1em]{}Upon successful authentication of Drone \texttt{A}, Drone \texttt{B} recovers the secret challenge $C_\emph{AB}^*$ by $C_\emph{AB}^*= X^*_A \oplus \mathcal{H}(s_\emph{BA},1)$. Since the shared secret $s_\emph{BA}$ is used for generating the OTP key, it cannot be reused in order to prevent ``two-time pad problem’’\footnote{In two-time pad attack, the same key $k$ is reused to OTP encrypt two different plaintexts, $p$ and $q$, then the ciphertexts $p \oplus k$ and $q \oplus k$ can be XORed together to recover $p \oplus q$.}~\cite{mason2006natural}. To refresh $s_\emph{BA}$, Drone \texttt{B} generates a new challenge $C_\emph{BA}^*$ to produce a response $R_\emph{BA}^*$ from its PUF. This new response is used to update $s_\emph{BG}$ to $s_\emph{BG}^*=\mathcal{H}(R_\emph{BA}^* \oplus n_B, \emph{ID}_G)$, and in turn, the shared secret $s_\emph{BA}$ to $s_\emph{BA}^*=\mathcal{H}(s_\emph{BG}^*, \emph{ID}_A)$ for the next session. \\
    \makebox[1em]{}For Drone \texttt{A} to generate the session key, it needs to have the updated shared secret $s_\emph{BA}^*$ generated by Drone \texttt{B}. To allow Drone \texttt{A} to recover this updated shared secret, Drone \texttt{B} encrypts $s_\emph{BA}^*$ into $X^*_B = s_\emph{BA}^* \oplus \mathcal{H}(s^\prime_\emph{BA},2)$ using $\mathcal{H}(s^\prime_\emph{BA}, 2)$ as the OTP key. Drone \texttt{B} then reciprocates the authentication by generating a credential $\mathscr{C}_B = \mathcal{H}(X^*_B \oplus s_\emph{BA}, C_\emph{AB}^*)$ and sends the message $M_2 = \langle X^*_B, \mathscr{C}_B \rangle$ to Drone \texttt{A}. \\
    \makebox[1em]{}Meantime, Drone \texttt{B} computes the session key $\emph{sk}=\mathcal{H}(C_\emph{AB}^*, s_\emph{BA}^*)$ by hashing the successfully recovered new challenge $C_\emph{AB}^*$ generated by Drone \texttt{A} and its generated shared secret $s_\emph{BA}^*$. Finally, $C_\emph{BA}$ is updated to the new challenge $C_\emph{BA}^*$ upon successful encrypted communication using the session key $\emph{sk}$ or before the current session ends. 
    
    \item \textbf{Drone \texttt{A} completes the authentication and derives the session key}:\\
    \makebox[1em]{}Upon receiving $M_2$, Drone \texttt{A} computes $\mathscr{C}^\prime_B = \mathcal{H}(X^*_B  \oplus s^\prime_\emph{BA}, C_\emph{AB}^*)$. If $\mathscr{C}^\prime_B=\mathscr{C}_B$, Drone \texttt{B} is authenticated since only Drone \texttt{B} who possesses the correct shared secret $s_\emph{BA}$ can successfully decrypt $X^*_A$ to recover $C_\emph{AB}^*$. \\
    \makebox[1em]{}Upon successful authentication of Drone \texttt{B}, \texttt{A} recovers the new shared secret $s_\emph{BA}^*$ by $X^*_B \oplus \mathcal{H}(s^\prime_\emph{BA}, 2)$ and applies $C_\emph{AB}^*$ to its embedded \emph{PUF} to obtain $R_\emph{AB}^*$. This response is used to update the D2G secret $s_\emph{AG}$ to $s_\emph{AG}^*=\mathcal{H}(R_{AB}^* \oplus n_A, \emph{ID}_G)$. The updated D2G secret is then used to update the OTP key for encrypting the new shared secret $s_\emph{BA}^*$ into $X_\emph{BA}^*= s_\emph{BA}^* \oplus \mathcal{H}(s_\emph{AG}^* \oplus n_A, \emph{ID}_B)$ for the next session. \\
     \makebox[1em]{}Drone \texttt{A} computes the session key $\emph{sk} = \mathcal{H}(C_\emph{AB}^*, s_\emph{BA}^*)$ for secure communication with Drone \texttt{B}. Upon successful encrypted communication using this session key or before the current session ends, Drone \texttt{A} updates $C_\emph{AB}$, $X_\emph{BA}$ to $C_\emph{AB}^*$ and $X_\emph{BA}^*$, respectively.
\end{enumerate}

\begin{figure}[!t]
\centering
\subfloat[]{
 \includegraphics[width=1\columnwidth]{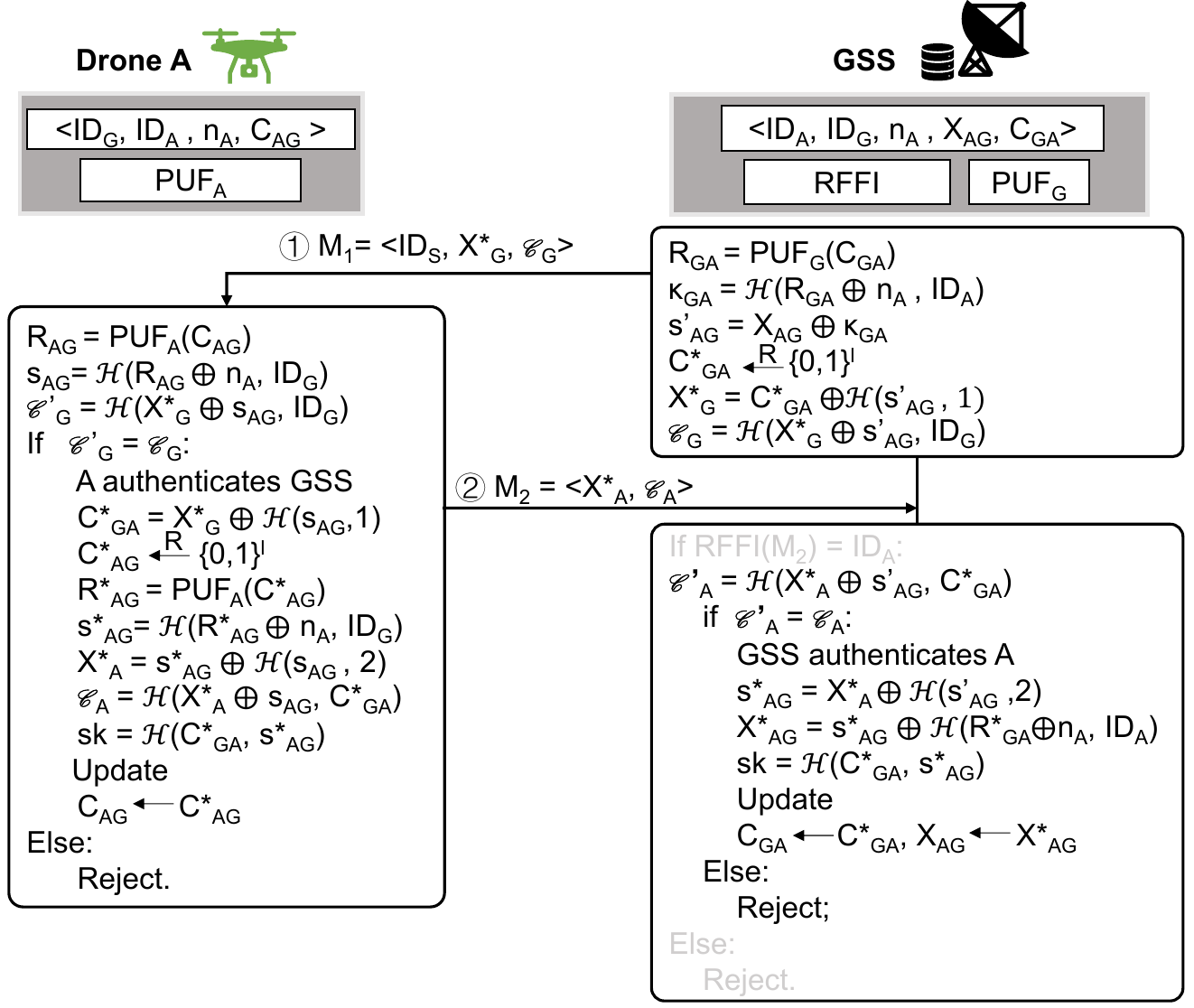}
 \label{fig:D2S_S_initiate}
}
\hfil
\subfloat[]{
 \includegraphics[width=1\columnwidth]{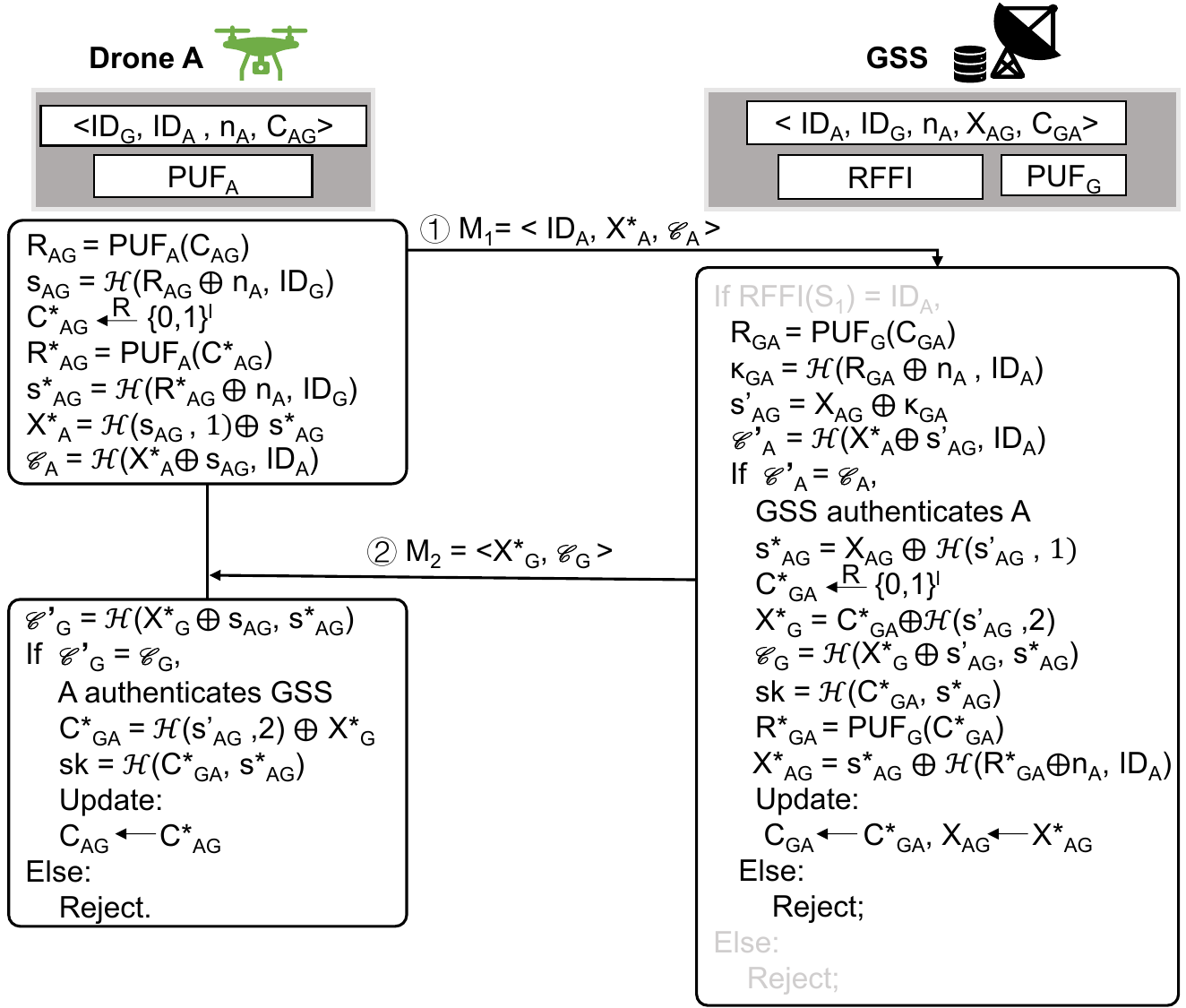}
 \label{fig:D2S_A_initiate}
}
\caption{D2G Mutual Authentication and Key Exchange phase. (a) Initiated by the \texttt{GSS}; (b) Initiated by Drone \texttt{A}.}
\label{fig:D2S}
\end{figure}

Fig. \ref{fig:D2D}(b) illustrates the protocol execution process when the D2D MAKE is initiated by Drone \texttt{B}. Although Drone \texttt{B} is enrolled earlier than Drone \texttt{A}, the OTA enrollment protocol described in Section \ref{sec:OTA enrolment} ensures that the secret $s_\emph{BA}$ generated from $s_\emph{BG}$ will be shared with Drone \texttt{A}, provided that Drone \texttt{A} is successfully enrolled with the \texttt{GSS}. Therefore, their mutual authentication follows similar steps, with the roles and order of prover and verifier swapped between Drones \texttt{A} and \texttt{B}. Drone \texttt{B} generates $s_\emph{BA}$ using its own PUF, while Drone \texttt{A} recovers $s^\prime_\emph{BA}$ from $X_B$. Both drones leverage the shared secret ($s_\emph{BA}=s^\prime_\emph{BA}$ for legitimate \texttt{A} and \texttt{B}) for mutual authentication and session key establishment. The two intermediate secrets, $s_\emph{BG}$ and $s_\emph{AG}$, are updated every round in the same way as in Fig. \ref{fig:D2D}(a), ensuring the freshness of $M_1$ and $M_2$ in every session to prevent replay attacks.

\subsubsection{D2G MAKE}
The D2G MAKE protocol follows a similar execution flow. As shown in Fig. \ref{fig:D2S}, either the \texttt{GSS} or the drone can initiate the MAKE process. Unlike the D2D MAKE protocol in Fig.~\ref{fig:D2D}, which uses the shared D2D secret $s_\emph{BA}$, the shared D2G secret $s_\emph{AG}$ ($s^\prime_\emph{AG}$) is used for mutual authentication between the \texttt{GSS} and Drone \texttt{A}. This shared key is updated by both parties in every session for enhanced security. Although the enrollment protocol in Fig. \ref{fig:enrollment_GSS} can also support D2G MAKE, it requires more interactions (and therefore delays) to achieve higher drone authentication accuracy at the \texttt{GSS} due to the adoption of RFFI. In addition, it consumes more resources on the drone due to the use of ECC encryption. While reasonable for enrollment, this may be excessive for frequent and time-sensitive D2G tasks after deployment. The proposed MAKE protocol, based on the shared D2G secret, is more efficient and lightweight. 

RFFI can also be utilized by the \texttt{GSS} to form a multi-factor authentication mechanism for authenticating drones during communication sessions (as shown in the grey text of Fig.~\ref{fig:D2S}). This method allows the drones to be continuously authenticated based on their RF signals, which provides higher security for the protected system compared to one-time authentication at the start of a session. 

\section{Security Analysis}
\label{sec:security}
In this section, we first analyze the resilience and immunity of both enrollment and MAKE phases against various types of attacks, and then provide an automated security analysis using the security verification tool ProVerif~\cite{ProVerifManual}.

\subsection{Threat Model and Assumptions}
The Dolev-Yao threat model~\cite{dolev1983security} is used for the security analysis. We assume the adversary can eavesdrop on all messages transmitted over a public channel, as well as modify or replay them within the network. The adversary may attempt to impersonate a legitimate drone or the \texttt{GSS}, launch a MITM attack, and is allowed to initiate an unbounded number of authentication requests. Furthermore, the adversary may physically capture a legitimate drone and attempt to clone the device or extract information from its memory. In the following analyses, for ease of exposition and without loss of generality, we use notations $s_\emph{BA}$ and $s_\emph{AG}$ to refer to the D2D and D2G secrets, respectively. The analyses are valid for other successfully enrolled drones besides Drones \texttt{A} and \texttt{B}.
\subsection{Informal Analysis}

\subsubsection{Replay, MITM and Impersonation Attacks}
These three attacks are widely discussed in IoD systems and are defined as follows: 
\begin{itemize}
\item Replay attacks: where an adversary retransmits previously intercepted messages to gain unauthorized access;
\item MITM attacks: where an adversary positions itself between two legitimate parties to intercept, read, and manipulate their communications;
\item Impersonation attacks: where an adversary pretends to be a legitimate drone or the \texttt{GSS} to deceive other nodes and steal sensitive data. 
\end{itemize}

All three attacks require sending replayed or altered/fake messages via the adversary’s transmitter. In the enrollment phase, thanks to the adoption of the RFFI system on the \texttt{GSS} to identify the message sender, the above three attacks can be easily detected by \texttt{GSS}. The message $M_2$ is also guaranteed to be authenticated and fresh per session due to the involvement of the D2G secret $s_\emph{AG}$ and the fresh nonce $n_A$ from Drone~\texttt{A} in its calculation, hence preventing MITM, impersonation, and replay attacks on Drone~\texttt{A}. In the MAKE phase, the authentication between drones fails if the adversary replays an outdated $M_1$ or $M_2$. These messages are refreshed per session due to the update of the shared D2D secret $s_\emph{BA}$ at the end of each MAKE session. Without knowing $s_\emph{BA}$, the adversary is not able to forge a valid $M_1$ or $M_2$, which helps prevent the MITM and impersonation attacks. 

\subsubsection{Cloning Attack}
A cloning attack refers to the adversary behavior of capturing a legitimate drone and attempting to replicate it, enabling the adversary to masquerade as the legitimate drone. Note that both drones and \texttt{GSSs} in our protocol are equipped with PUFs. Any physical, invasive attempts to probe or alter the monolithically integrated PUF circuit of the victim device will disrupt its sensitive physical characteristics -- such as variations in delay, capacitance, or initial memory state -- and consequently, permanently change or destroy its unique challenge-response mapping. Additionally, since the \texttt{GSS} authenticates drones with RFFI, cloning a drone necessitates replicating its transmitter as well. Similar to PUF, RFFI leverages hardware-induced imperfections from the transmitter’s manufacturing process, making the resulting signal extremely difficult, if not impossible, to perfectly replicate or clone in a practical sense.

\subsubsection{DoS Attack}
A DoS attack is a malicious attempt to disrupt the normal traffic of a targeted node, service, or network by overwhelming it with excessive traffic, rendering it unable to respond to legitimate requests. 
In the enrollment phase, the \texttt{GSS} employs RFFI to verify the sender’s legitimacy. If the message is identified as originating from an adversary, the \texttt{GSS} discards it immediately and halts further processing. In the MAKE phase, lightweight mutual authentication is performed at the beginning of each session, which effectively helps the receiver determine whether the message comes from a pre-enrolled, legitimate drone. RFFI can also be utilized for continuous authentication during the MAKE process for the D2G communication scenario. Malicious requests can be rejected promptly to prevent DoS attacks. 

\subsubsection{Mutual Authentication}
During the enrollment phase, the \texttt{GSS} authenticates drones by verifying their RFFs. Subsequently, Drone \texttt{A} authenticates the \texttt{GSS} by validating the correctness of $M_2$, which requires knowledge of the D2G secret $s_\emph{AG}$. This D2G secret can only be derived by a legitimate \texttt{GSS} with the correct private key $\emph{sk}_G$. In the MAKE phase, mutual authentication between Drone \texttt{B} (or \texttt{GSS}) and Drone \texttt{A} is achieved by verifying $\mathscr{C}_B$(or $\mathscr{C}_G$) and $\mathscr{C}_A$. A match between the prover-generated and verifier-recalculated versions of these credentials confirms that both parties share the same D2D secret $s_\emph{BA}$ or D2G secret $s_\emph{AG}$. These secrets could only be obtained upon successful OTA enrollment, thereby ensuring their authenticity. 

\subsubsection{PFS}
To achieve PFS, our proposed protocol eliminates the need for any long-term secrets. Specifically, all secrets in the protocol, including the D2G and D2D secrets or OTP keys, are derived from PUF responses $R_\emph{AG}$, $R_\emph{AB}$, or $R_\emph{BA}$. These PUF responses, serving as the root secrets of the system, are renewed in every session for the computation of $X^*_B$ or $X^*_G$ during the execution of the protocol. With knowledge of the current session root key $s_\emph{BA}^*$ and the previously collected $X^*_B$, an attacker can only recover $\mathcal{H}(s_\emph{BA},2)$. Due to the one-wayness of the cryptographic hash function, the previous session root key $s_\emph{BA}$ cannot be deduced by a PPT adversary. Therefore, the exposure of current root keys will not help the adversary to compromise the secrecy of session keys established before the exposure. In addition, strong PUFs are designed to produce responses that are statistically independent and unpredictable for an unknown challenge, even if a large number of other CRPs are known. Consequently, the adversary cannot infer previous unexposed responses even if subsequent ones are exposed or compromised, making it impossible to deduce the previous session key $\emph{sk}$. Thus, even if the shared secret or the current session key is leaked, the previous ciphertext communication records will still be secure.

\subsection{Formal Security Verification Using ProVerif}
ProVerif~\cite{ProVerifManual} is a formal analysis tool of security protocols. It can be used to automatically verify whether a protocol meets specified security requirements, including mutual authenticity, key secrecy, resistance against replay and MITM attacks, etc., using a process calculus language under the Dolev-Yao~\cite{dolev1983security} security model. Standard cryptographic operations, such as encryption or signature algorithms, are modeled as constructor-destructor pairs, where the constructor represents the operations (e.g., encryption or signing) and the destructor represents their inverse (e.g., decryption or signature validation algorithms). These are treated as black-box functions, abstracting away their internal implementations. 

PUF and RFFI are two essential operations in our proposed protocol. As unconventional security primitives, neither PUF nor RFFI has standard ProVerif constructors and destructors. To facilitate ProVerif analysis, we create behavior models for PUF and RFFI that simulate their respective essential properties without specifying internal implementations. Specifically, we model the PUF as a private one-way function that allows only the legitimate entity to generate a response given the challenge, making it impossible to deduce the challenge based on the response. For RFFI, we model its transmitter traceability property with the following two constructors and a destructor: 
\begin{itemize}
\item A constructor $\textbf{Ex\_A}(nonce, \emph{ID}_A$): a private function that extracts a distinct variation ($\emph{RFF}_A$) as RFF from identity $\emph{ID}_A$. As an adversary who intercepts a message can only replay the message through her/his own transmitter, the replayed message will contain the adversary's distinct RFF, not the legitimate device's. An RFFI system at the receiver can detect this mismatch, identifying the transmission as originating from an unauthorized device. To model the subtle, random deviations in the RF waveforms of different transmitters sending the same message, a nonce is incorporated in this constructor to represent the freshness (uniqueness) of their RF features that the RFFI can recover.
\item A constructor $\textbf{P} (payload, \emph{RFF}_A)$: a public function that models the distinct fingerprint $\emph{RFF}_A$ of $\emph{ID}_A$ carried in a message with the \textit{payload}. 
\item A destructor $\textbf{RFFI}()$: a public function that can recover the original payload, the nonce, and the transmitter identity (ID), defined as: $\textbf{RFFI}(\textbf{P}(payload,\emph{RFF}_A))=(payload, nonce, \emph{ID}_A)$, thereby identifying the source transmitter of the message.
\end{itemize}

\begin{figure}[!t]
    \centering
    \includegraphics[width=1\linewidth]{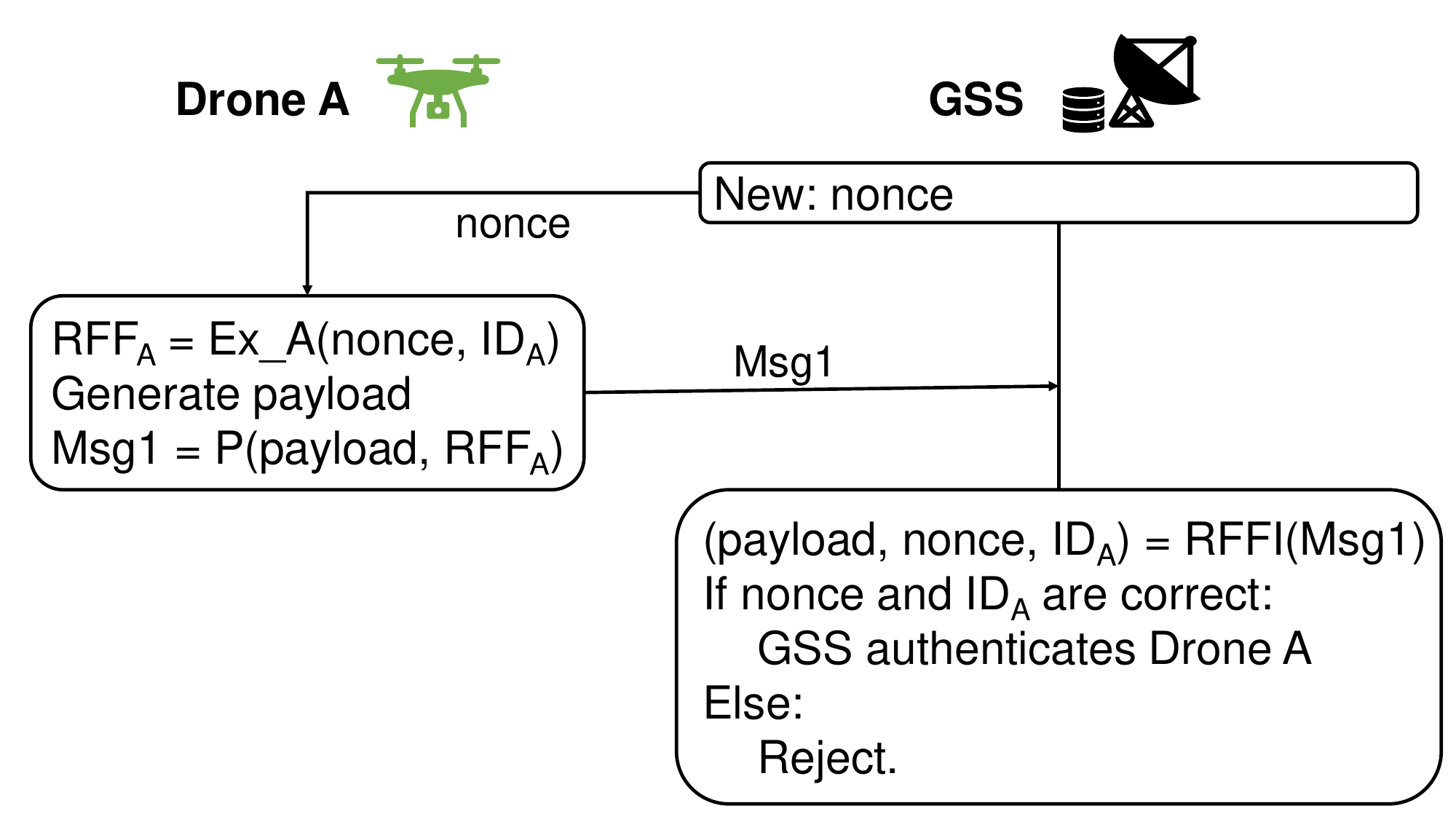}
    \caption{RFFI model in ProVerif.}
    \label{fig:RFFI_ProVerif}
\end{figure}

As illustrated in Fig. \ref{fig:RFFI_ProVerif}, Drone \texttt{A} generates $\emph{Msg1}=\textbf{P}(payload, \textbf{Ex\_A}(nonce, \emph{ID}_A))$ to simulate the radio frequency signal that carries the packet and exhibits its RFF ($\emph{RFF}_A$), while the \texttt{GSS} uses $\textbf{RFFI}(\emph{Msg1})=\textbf{RFFI}(\textbf{P}(payload, \textbf{Ex\_A}(nonce, \emph{ID}_A)))=(payload, nonce, \emph{ID}_A)$ to extract the payload, the nonce and corresponding ID. When the nonce and ID are correct, the \texttt{GSS} authenticates Drone \texttt{A}.

\begin{figure}[!t]
\centering
\subfloat[]{
 \includegraphics[width=1\columnwidth]{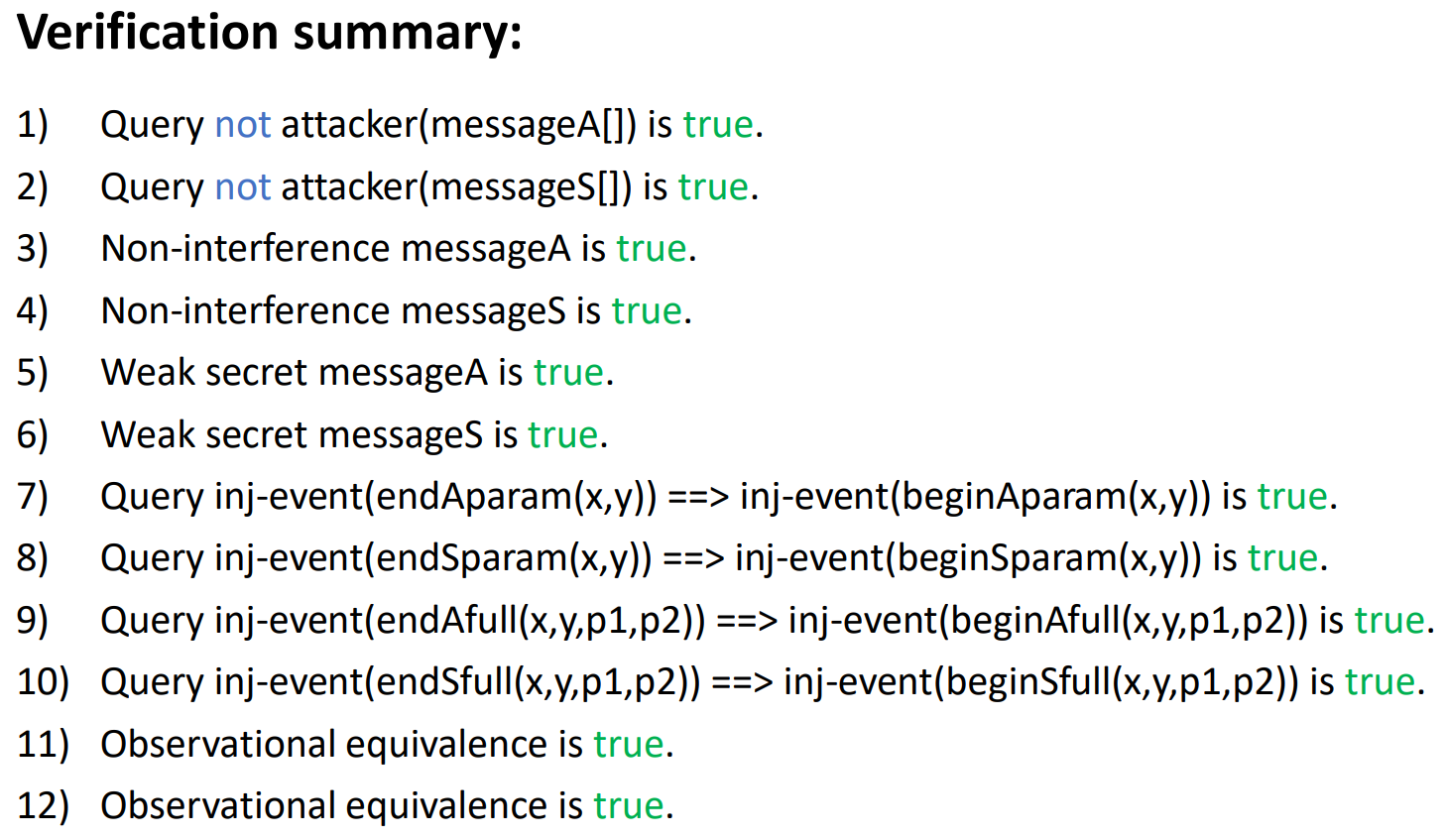}
 \label{fig:PV_enroll}
}
\hfil
\subfloat[]{
 \includegraphics[width=1\columnwidth]{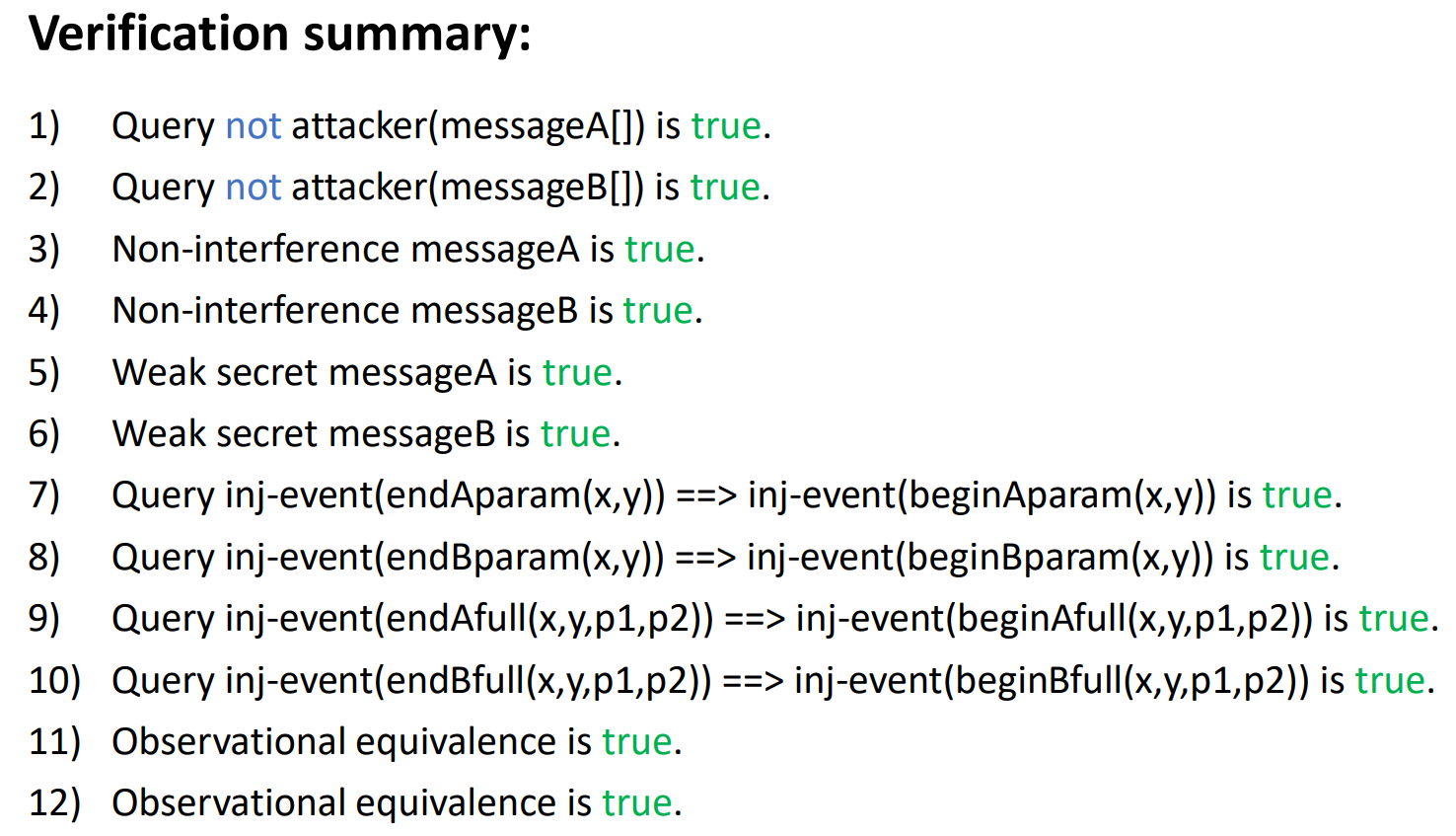}
 \label{fig:PV_MAKE}
}
\caption{ProVerif output summary of (a) Enrollment phase and (b) MAKE phase. (1-2): Reachability proof; (3-4): Strong secrecy proof; (5-6): The absence of off-line guessing attacks proof; (7-10): Mutual authentication proof; (11-12): OE between processes that differ only by terms and real-or-random secrecy proof. A ``true" output indicates the proof is correct.}
\label{fig:ProVerif}
\end{figure}

Based on the formal modeling described above, the proposed protocol was implemented in ProVerif and subjected to comprehensive verification. The simulation covered an unbounded number of protocol sessions and explored the entire message space between participating entities. The security evaluation focused on three critical aspects: 
\begin{itemize}
    \item Mutual authentication: ensuring valid and secure mutual authentication between communicating parties while demonstrating resistance against replay and MITM attacks; 
    \item Reachability and secrecy: assessing the reachability and secrecy of shared secrets, i.e., $s_\emph{AG}$ and $s_\emph{BA}$, in the enrollment phase and the session key $\emph{sk}$ in the MAKE phase;
    \item Observation equivalence (OE): evaluating strong secrecy, real-or-random secrecy, and resistance to off-line guessing attacks. This analysis determines whether an adversary can detect secret change, differentiate between actual secrets and random values, or correctly guess the secrets. A detailed discussion of the correspondence assertions used in this analysis can be found in~\cite{zheng2022puf}. 
\end{itemize}
Fig. \ref{fig:ProVerif} presents the automatic verification summary of our protocol on both the enrollment and MAKE phases, confirming that the proposed protocol satisfies all three security objectives.

\section{Performance Analysis}
\label{sec:performance}
In this section, we compare the proposed work with representative D2D~\cite{pu2022lightweight, lounis2022d2d, karmakar2023puf} and D2G~\cite{sen2025securing,zhou2025radio} protocols for IoD in terms of security features, computational complexity, as well as the communication and storage costs.

\subsection{Comparison of Security Properties}
Table \ref{tab:securitycomparison} presents a comparison of the protocols across various security features, including mutual authentication, key agreement, PFS, the absence of an intermediary (Two-Party Authenticated Key), resistance against DoS attacks (DOS Resistance), supporting OTA enrollment, and no exploitable secret is stored on drones (secret-free onboard storage). 
\begin{table*}[!t]
\centering
\caption{Comparison of Security Features}
\begin{threeparttable}
\begin{tabular}{lccccccc}
\toprule
\textbf{Protocols} & \textbf{MA}\tnote{1} & \textbf{KA}\tnote{2}& \textbf{PFS} &\textbf{2PAKE }\tnote{3} & \textbf{DoS Resistance} & \textbf{OTA Enrollment} & \textbf{Secret-Free Onboard Storage}\\ \hline
Pu \textit{et al.} \cite{pu2022lightweight} (D2D) & \checkmark  & \checkmark  &\checkmark & $\times$ & \checkmark & $\times$ & \checkmark \\
Lounis \textit{et al.} \cite{lounis2022d2d} (D2D) & \checkmark  & \checkmark  &\checkmark & \checkmark & $\times$ & $\times$ & $\times$ \\ 
Karmakar \textit{et al.} \cite{karmakar2023puf} (D2D) & \checkmark  & \checkmark  &\checkmark & $\times$ &\checkmark & $\times$ & $\times$ \\ 
Sen \textit{et al.} \cite{sen2025securing} (D2G) & \checkmark  & \checkmark  &\checkmark & $\times$ &\checkmark & $\times$ & \checkmark \\ 
Zhou \textit{et al.} \cite{zhou2025radio} (D2G) & \checkmark  & \checkmark  & $\times$ & \checkmark &\checkmark & $\times$ & $\times$ \\ 
This work (D2D\&D2G) & \checkmark & \checkmark & \checkmark & \checkmark & \checkmark & \checkmark & \checkmark \\ \bottomrule
\end{tabular}%
\begin{tablenotes}
    \small
\item [1] Mutual Authentication; $^2$Key Agreement; $^3$Two-Party Authenticated Key Exchange.
\end{tablenotes}
\end{threeparttable}
\label{tab:securitycomparison}
\end{table*}

All evaluated protocols can meet the basic requirements of mutual authentication and key agreement. The protocol~\cite{zhou2025radio} cannot achieve PFS, and the three protocols, \cite{pu2022lightweight, karmakar2023puf}. and~\cite{sen2025securing}, require a third party's participation for mutual authentication. Since the drone in the protocol of~\cite{lounis2022d2d} computes and sends back a message upon receiving a syntactically correct message without first authenticating the sender, the protocol is vulnerable to DoS attacks. An attacker can exhaust the drone's battery by sending a flood of forged requests. The protocols, \cite{lounis2022d2d} and~\cite{zhou2025radio}, store unprotected secret data directly in memory. If a single drone is captured, lost, or destroyed in these systems, an adversary can discover meaningful sensitive information from its storage. The security of the authentication protocol~\cite{karmakar2023puf} rests on the assumption that the mechanism for splitting the secret pseudo identity into three parts is kept private to both the drones and the ground station. If this splitting mechanism of the protocol is kept confidential, then it violates Kerckhoffs's Principle. Therefore, the splitting mechanism must be hardcoded or directly stored; securely storing it otherwise could pose a significant burden for resource-constrained drones. In contrast, our protocol is the only one that achieves secure OTA enrollment and fulfills all the listed desirable security properties. It can perform drone enrollment with the target \texttt{GSS} without requiring a secure environment. This capability provides significant convenience, flexibility, and scalability to the IoD system by allowing drones to be provisioned and authenticated remotely, which enables them to be deployed rapidly to exchange information.

\subsection{Comparison of Computational Complexity}

This section compares the computational complexities of these protocols based on the operations performed during the authentication phase. To ensure a fair comparison and minimize bias that may arise from different implementation platforms, the analysis focuses on essential operations required for each participating party. Trivial operations, such as concatenation and exclusive OR (XOR), are omitted because their impact on the relative accuracy of the comparison is negligible. The comparison primarily considers functions of cryptographic modules (e.g., PUF, Hash) unless a protocol explicitly specifies otherwise. This approach is taken because various specific implementations can be employed as long as they meet the functional requirements and cost constraints of the system. Furthermore, PUF reliability-enhancement techniques (e.g., majority voting, error correction codes) are not included for any protocol. This maintains a consistent baseline for comparison and prevents the analysis from favoring specific PUF implementations.

The execution times for one-round operations were measured. To ensure a fair comparison, specific instantiations of cryptographic primitives and the PUF were implemented in Python on an Avnet Ultra96-V2 board. The measured execution time for a 256-bit input is reported in Table \ref{tab:ETULTRA96}, detailing the following components and their respective time notations:
\begin{itemize}
\item $T_\emph{PUF}$: PUF, instantiated as an Arbiter PUF.
\item $T_H$: Hash function, instantiated as SHA-256.
\item $T_\emph{RS}$: Random shuffling (replicated from \cite{pu2022lightweight}).
\item $T_\emph{ME}$: Modular Exponentiation.
\item $T_\emph{Sig}$: Digital Signature Generation, instantiated as ECDSA (Elliptic Curve Digital Signature Algorithm). 
\item $T_\emph{Verf}$: Signature Verification.
\item $T_\emph{MAC}$: Message Authentication Code (MAC) operation, instantiated as    
Hash-based MAC (HMAC).    
\end{itemize}
\begin{table}[!t]
\caption{Execution Time on ULTRA96-V2}
\label{tab:ETULTRA96}
\centering
\begin{threeparttable}
\begin{tabular}{ccc}
\toprule
\textbf{Operation} & \textbf{Symbol} & \textbf{Execution Time (ms)} \\ \midrule
Arbiter PUF & $T_\emph{PUF}^*$ & 0.5658 \\ 
SHA256 & $T_H$ & 0.0066 \\ 
Random Shuffling & $T_{\emph{RS}}$ & 0.3049 \\ 
Modular Exponential & $T_{\emph{ME}}$ & 1.8458 \\ 
ECDSA Digital Signature Generation & $T_{\emph{Sig}}$ & 19.7414 \\ 
ECDSA Signature Verification & $T_{\emph{Verf}}$ & 38.8412 \\ 
HMAC & $T_{\emph{MAC}}$ & 0.0318 \\ 
\bottomrule
\end{tabular}
\begin{tablenotes}
    \small
    \item [*] The execution time for generating a 256-bit response from the 256-stage Arbiter PUF. 
\end{tablenotes}
\end{threeparttable}
\end{table}

The computational complexity comparison presented in Table \ref{tab:computational_complexity} reveals a clear hierarchy. The protocol proposed in~\cite{zhou2025radio} incurs the highest computational overhead, a burden significantly greater than other schemes, due to its reliance on digital signatures and RFF extraction during the authentication process. In contrast, our solution demonstrates a lighter computational footprint compared to those in~\cite{pu2022lightweight} and~\cite{lounis2022d2d}, and it is on par with the protocols introduced in~\cite{karmakar2023puf} and~\cite{sen2025securing}. However, the scheme from ~\cite{karmakar2023puf} necessitates third-party involvement for mutual authentication between devices, which increases the number of communication rounds and offers fewer security features than our solution. Similarly, the scheme in ~\cite{sen2025securing} has a critical vulnerability: it stores drones’ raw CRPs in the \texttt{GSS}, making it susceptible to the stolen verifier attack. It also requires third-party involvement. Overall, our proposed protocol provides a broader array of security features and ensures superior efficiency without requiring a trusted third party. It consistently outperforms existing solutions by offering low computational cost, zero local key storage, and excellent flexibility.  
\begin{table*}[t]
\centering
\caption{Comparison of Computational Complexity}
\label{tab:computational_complexity}
\begin{threeparttable}
\begin{tabular}{lcccc}
\toprule
\textbf{Protocols} & \textbf{Drone A} & \textbf{Drone B/GSS} & \textbf{Third Party} & \textbf{Total} \\ \hline
Pu \textit{et al.} \cite{pu2022lightweight} (D2D) & \makecell{3$T_H$+5$T_\emph{RS}$+4$T_\emph{MAC}$+$T_\emph{PUF}$} & \makecell{3$T_H$+3$T_\emph{RS}$+2$T_\emph{MAC}$+$T_\emph{PUF}$} & \makecell{2$T_H$+6$T_\emph{RS}$+6$T_\emph{MAC}$} & \makecell{8$T_H$+14$T_\emph{RS}$+12$T_\emph{MAC}$+2$T_\emph{PUF}$\\$\approx$ 5.84ms}  \\
Lounis \textit{et al.} \cite{lounis2022d2d} (D2D) & 6$T_\emph{PUF}$+8$T_H$ & 6$T_\emph{PUF}$+12$T_H$ & / & 12$T_\emph{PUF}$+20$T_H\approx 6.92ms$ \\ 
Karmakar \textit{et al.}  \cite{karmakar2023puf} (D2D) & 2$T_\emph{PUF}$+5$T_H$ & 2$T_\emph{PUF}$+5$T_H$ & 9$T_H$ & 4$T_\emph{PUF}$+19$T_H \approx 2.39ms$ \\
Sen \textit{et al.} \cite{sen2025securing} (D2G) & 3$T_\emph{PUF}$+2$T_H$ & 3$T_\emph{PUF}$+2$T_H$ & 4$T_H$ & 6$T_\emph{PUF}$+8$T_H \approx 3.45ms$ \\
Zhou \textit{et al.} \cite{zhou2025radio} (D2G) & \makecell{8$T_\emph{ME}$+$T_\emph{Sig}$ +2$T_\emph{MAC}$} & \makecell{$T_\emph{Verf}$+8$T_\emph{ME}$+2$T_\emph{MAC}$+$T_\emph{RFF}$} & / & \makecell{$T_\emph{Verf}$+16$T_\emph{ME}$+$T_\emph{Sig}$+4$T_\emph{MAC}$+$T_\emph{RFF}$ \\$\approx 88.24ms + T_\emph{RFF}^*$}  \\ 
This work (D2D) & 2$T_\emph{PUF}$+9$T_H$ & 2$T_\emph{PUF}$+9$T_H$ & / & 4$T_\emph{PUF}$+18$T_H \approx 2.38ms$\\
This work (D2G) & 2$T_\emph{PUF}$+7$T_H$ & 2$T_\emph{PUF}$+7$T_H$ & / & 4$T_\emph{PUF}$+14$T_H \approx 2.36ms$ \\ \bottomrule
\end{tabular}
\begin{tablenotes}
    \small
    \item[*] The execution time required for RFF extraction is not quantified due to implementation uncertainty.
\end{tablenotes}
\end{threeparttable}
\end{table*}

\subsection{Comparison of Communication and Storage Costs}

For a fair comparison of communication and storage costs, we standardized the sizes of several security parameters. Device identities and timestamps were set to 32 bits. Hash outputs, random nonces, keys, elliptic curve points, and challenges and responses of PUFs were all set to 256 bits. 

As illustrated in Table \ref{tab:communication_comparison}, protocols from~\cite{pu2022lightweight,karmakar2023puf, sen2025securing} show slightly lower storage costs on the drones. However, this advantage is achieved by offloading some storage burden to third parties, which consequently results in significantly higher communication overhead. The protocol detailed in~\cite{lounis2022d2d} supports direct D2D mutual authentication, yet its communication overhead also remains substantially higher than the aforementioned schemes. A related work in~\cite{zhou2025radio}, similar to our proposed D2G protocol, achieves direct mutual authentication using RFF for drone authentication. However, that approach requires more protocol rounds and the transmission of more bits. In contrast, our proposed approach achieves mutual authentication in both D2D and D2G scenarios with only two message exchanges. Our design minimizes bandwidth consumption (approximately 1000 bits) and demands relatively low storage requirements on the drone. 
\begin{table*}[!t]
\centering
\caption{Comparison of Communication and Storage Costs}
\begin{threeparttable}
\begin{tabular}{lccccccc}
\toprule
\multirow{2}{*}{\textbf{Protocols}} & \multirow{2}{*}{\textbf{Message Exchanges}} & \multicolumn{4}{c}{\textbf{Communication Cost}} & \multicolumn{2}{c}{\textbf{Storage Cost (per pair)}} \\
\cmidrule(lr){3-6} \cmidrule(lr){7-8}
 & & \textbf{Drone A} & \textbf{Drone B/GSS} & \textbf{Third Party} &  \textbf{Total Cost}  & \textbf{Drone A} & \textbf{Drone B/GSS} \\
\midrule
Pu \textit{et al.} \cite{pu2022lightweight}(D2D) & 6 & 1536 bits& 768 bits & 2048 bits  & 4352 bits & 288 bits & 288 bits \\
Lounis \textit{et al.} \cite{lounis2022d2d}(D2D) & 5 & 3776 bits & 3200 bits & / & 6976 bits & 1280 bits & 1280 bits \\
Karmakar \textit{et al.} \cite{karmakar2023puf} (D2D) & 3 & 2336 bits & 1536 bits & 1536 bits & 5408 bits & 544 bits & 544 bits \\
Sen \textit{et al.} \cite{sen2025securing} (D2G) & 8 & 832 bits & 832 bits & 3072 bits & 4736 bits & 288 bits & 288 bits \\
Zhou \textit{et al.} \cite{zhou2025radio} (D2G) & 4 & 1600 bits & 1312 bits & / & 2912 bits & 512 bits & $S_\emph{RFF}^*$ +288 bits \\
This work (D2D) & 2 & 544 bits & 512 bits & / & 1056 bits & 864 bits & 608 bits \\
This work (D2G) & 2 & 544 bits & 512 bits & / & 1056 bits & 832 bits & $S_\emph{RFF}^*$ +576 bits \\
\bottomrule
\end{tabular}
\begin{tablenotes}
    \item[*] The model size of the RFF authentication model.
\end{tablenotes}
\end{threeparttable}
\label{tab:communication_comparison}
\end{table*}

In summary, the comparisons detailed in Tables \ref{tab:securitycomparison}, \ref{tab:computational_complexity}, and \ref{tab:communication_comparison} demonstrate that our proposed protocol achieves scalability, fulfills a wider range of security requirements, and incurs relatively low computational complexity, communication overhead, and storage costs.

\section{Conclusion}
\label{sec:conclusion}

This paper introduces a lightweight authentication protocol designed specifically for secure data transmission in IoD scenarios. The protocol leverages the unique physical properties of both PUF and RFFI. A key feature is its ability to ensure direct mutual authentication for both D2D and D2G communications. By using PUF and OTP encryption, the protocol allows resource-constrained drones to operate without storing any secrets locally. The effective utilization of RFF facilitates a streamlined, OTA enrollment process at the \texttt{GSS}, boosting scalability and feasibility. The system also guarantees PFS through dynamic, per-session updating of root keys, which eliminates long-term secret dependencies. The security of the proposed protocol has been rigorously validated via both informal analysis and formal verification using the ProVerif tool under the robust Dolev-Yao threat model. The result is a more efficient solution that demonstrates significantly lower communication cost and computational complexity compared to existing IoD authentication protocols.



%






%


\bibliographystyle{IEEEtran}

\bibliography{IEEEabrv,references}

\end{document}